\documentclass{aa} 

\usepackage{graphicx} 
\usepackage{subcaption} 
\usepackage{txfonts} 
\usepackage{xspace} 
\usepackage{soul} 
\soulregister\ref7 
\soulregister\cite7
\soulregister\citet7
\soulregister\citep7
\soulregister\mathcal7
\soulregister\mathrm7
\soulregister\texttt7
\usepackage{xcolor} 

\usepackage[
  bookmarksopen=false,
  bookmarksnumbered=true,
  breaklinks=true,
  colorlinks=true,
  linkcolor=blue,
  citecolor=blue
]{hyperref} 

\makeatletter 
\renewcommand*{\corrauth}[1]{\thanks{Corresponding author:\\{\aa@emailfont #1}}}
\makeatother 

\begin{document}

\title{A self-consistent orbital architecture for GG Tau A}

\subtitle{I. Simultaneous orbital fitting of the hierarchical triple}

\author{
A. Lacquement \inst{1}\corrauth{antoine.lacquement@univ-grenoble-alpes.fr}
\and  H. Beust \inst{1}\email{herve.beust@univ-grenoble-alpes.fr}
\and G. Duch\^ene \inst{1,2}\email{gaspard.duchene@univ-grenoble-alpes.fr}
\and C. Lawlor \inst{3}\email{c.lawlor13@universityofgalway.ie}
\and C. Ginski \inst{3}\email{christian.ginski@universityofgalway.ie}
\and E. Di Folco \inst{4}\email{emmanuel.di-folco@u-bordeaux.fr}
\and A. Dutrey \inst{4}\email{anne.dutrey@u-bordeaux.fr}
}

\institute{
$^{1}$Université Grenoble Alpes, CNRS, IPAG, 38000 Grenoble, France\\
$^{2}$Department of Astronomy, University of California, Berkeley, CA 94720, USA\\
$^{3}$School of Natural Sciences, Center for Astronomy, University of Galway, Galway H91 CF50, Ireland\\
$^{4}$Université de Bordeaux, CNRS, LAB, 33615 Pessac, France\\
}

\date{Received 22 July 2026 / Accepted 25 August 2026}

\abstract
{
GG~Tau~$A$ is a young triple system consisting of the close pair, $Ab_1$--$Ab_2$, and a third component, $Aa$, surrounded by a massive circumtriple disk whose large inner cavity is difficult to explain in a purely binary framework. Recent astrometric studies have constrained the inner and wide orbits separately. However, only a joint treatment can provide a self-consistent description of the system and constrain the individual stellar masses.
}
{
We aim to determine the orbital architecture and individual stellar masses of GG~Tau~$A$ and assess how the available astrometric and disk-based constraints restrict the range of admissible solutions.
}
{
We performed a joint fit using \texttt{Oracle}, developed specifically for hierarchical stellar systems. All astrometric measurements were placed in a common reference frame, since the historical wide-orbit astrometry is given relative to the unresolved photocenter of the $Ab$ subsystem. The fit included one new wide-orbit astrometric epoch and a prior on the total stellar mass derived from disk kinematics. We then applied, in post-processing, an additional geometrical constraint based on the observed center of the circumtriple disk.
}
{
The fit yields orbital solutions compatible with the available astrometric and disk-based constraints and provides estimates of the individual stellar masses. The additional wide-orbit epoch only marginally reduces the range of admissible solutions. By contrast, the disk-center constraint leaves the favored orbital architectures largely unchanged but significantly tightens the stellar-mass partition. This yields posterior masses of $0.521^{+0.069}_{-0.051}$, $0.106^{+0.017}_{-0.013}$, and $0.79^{+0.10}_{-0.10}$~$M_\odot$ for $Ab_1$, $Ab_2$, and $Aa$, respectively. The reported values are posterior medians with 16th--84th percentile intervals.
}
{
A joint treatment of the two orbital levels is required to recover a physically meaningful architecture and constrain the individual stellar masses. The resulting solutions provide a basis for future dynamical modeling of the circumtriple disk and for testing whether the observed cavity can further constrain the system architecture.
}


\maketitle
\nolinenumbers

\section{Introduction}
\label{introduction}

Stellar multiple systems are a common outcome of star formation \citep{DucheneKraus2013}. For long-term stability, they are often organized in a hierarchical architecture \citep{Tokovinin2021}. In young systems, this architecture can strongly influence the evolution of circumstellar material, as disks are frequently still present during the early stages of evolution \citep{WilliamsCieza2011}. Hierarchical multiple systems therefore provide valuable laboratories for investigating the interplay between stellar dynamics and disk evolution \citep{Offner2023}. Mutual inclinations between orbital planes can drive secular evolution \citep{Naoz2013,Naoz2016}, while the hierarchical architecture can truncate disks, drive precession or polar alignment, and modulate accretion flows \citep{ArtymowiczLubow1994,Ceppi2022,Ceppi2023,Lepp2025}. Constraining the orbits is therefore a necessary step toward understanding both the stellar dynamics and the evolution of circumstellar material in young multiple systems.

GG~Tau~$A$ is one of the clearest benchmark systems for studying these questions. Located at a distance of approximately 145~pc \citep{Galli2019}, it is a very young system with an age generally estimated between 1 and 4~Myr \citep{White1999,Hartigan2003,KrausHillenbrand2009}. Early high-angular-resolution observations established its multiple nature and laid the observational foundations for the architecture inferred today \citep{Leinert1993,White1999,Hartigan2003}. Millimeter interferometric data subsequently revealed a dense circumbinary disk, together with a large central cavity, around what was still interpreted as a binary \citep{Dutrey1994,Guilloteau1999}. Additional scattered-light and near-infrared polarimetric observations resolved the inner edge of the dust disk, located at roughly 180~au, and revealed that its apparent morphology depends on wavelength \citep{Silber2000,Krist2002,Duchene2004}. Radio and millimeter observations later showed that millimeter-sized grains are concentrated farther out, within a relatively narrow ring extending from approximately 200 to 250~au and that this disk is among the most massive in Taurus, with a mass of 0.15~$M_\odot$ \citep{Guilloteau1999,Andrews2014,Dutrey2014,Phuong2020}. Although the large-scale emission remains broadly consistent with a nearly axisymmetric ring centered on a single geometrical position, higher-resolution observations also reveal localized azimuthal substructures together with streamer-like features within the cavity \citep{Beck2012,Phuong2018,Phuong2020,Keppler2020}.

The inner edge of the ring long remained difficult to reconcile with a purely binary interpretation of the system, as already pointed out by \citet{McCabe2002}. Using the astrometry available at the time, \citet{BeustDutrey2005} showed that the best-fit binary orbit would carve an inner cavity roughly half the observed size, whereas an orbit wide enough to reproduce the cavity remained only marginally compatible with the data. \citet{Koehler2011} reached a similar conclusion and further showed that orbits wide enough to explain the cavity had to be misaligned with respect to the disk. This difficulty was subsequently explored through hydrodynamical simulations of the disk \citep{NelsonMarzari2016,Cazzoletti2017,Aly2018,Keppler2020}. In particular, \citet{Keppler2020} report that a coplanar binary with a semimajor axis of approximately 35~au could qualitatively reproduce a large cavity together with several observed disk substructures after long-term hydrodynamical evolution. The picture nevertheless remains uncertain, and even recent models still struggle to reproduce simultaneously the stellar astrometry, cavity size, and observed disk morphology over timescales comparable to the age of the system \citep{Toci2024}.

A major change in this picture came from \citet{DiFolco2014}, who resolved GG~Tau~$Ab$ into the close pair $Ab_1$--$Ab_2$ and showed that GG~Tau~$A$ is a hierarchical triple rather than a simple binary. This discovery fundamentally changed the interpretation of the system. In particular, the outer ring can no longer be related solely to the $Aa$--$Ab$ orbit but must instead reflect the coupled influence of the inner subsystem and the wide orbit of $Aa$ around the center of mass of the close pair, denoted $Ab_\mathrm{cm}$.

Recent work has considerably improved the orbital picture, while also highlighting its current limitations. \citet{Duchene2024} report the first robust orbital solution for the inner pair $Ab_2$--$Ab_1$, whereas \citet{Toci2024} constrain the wide orbit using astrometry of the photocenter of the close pair, denoted as $Ab_\mathrm{pc}$. These two solutions cannot be combined directly because they refer to different physical quantities. The relation between $Ab_\mathrm{pc}$ and $Ab_\mathrm{cm}$ is itself set by the instantaneous configuration of the inner binary. Separate fits therefore do not provide a fully self-consistent description of the system. The inner orbit determines the offset between $Ab_\mathrm{pc}$ and $Ab_\mathrm{cm}$, while the wide orbit must be defined with respect to $Ab_\mathrm{cm}$ itself. A consistent interpretation thus requires the inner and wide orbits to be solved simultaneously. This joint treatment is also essential for the mass analysis itself. Because the available observables are purely relative astrometric measurements, they do not by themselves constrain how the total mass is partitioned among the three stars. Only the simultaneous modeling of the coupled inner and wide orbits makes it possible to recover direct constraints on the individual stellar masses. An analogous approach has already been applied to the T~Tau system by \citet{Duchene2006} and was later revisited with a more extensive dataset by \citet{Schaefer2020}. In the case of GG~Tau~$A$, one additional astrometric epoch of the wide subsystem has also become available since the publication of these studies, extending the observational baseline.

In this paper, we present a simultaneous hierarchical fit of the $Ab_2$--$Ab_1$ and $Aa$--$Ab_\mathrm{cm}$ orbits within a unified model. Our analysis incorporates the new wide-orbit measurement and converts the historical $Aa$--$Ab_\mathrm{pc}$ astrometry into $Aa$--$Ab_1$ measurements using an updated solution for the inner orbit. This construction yields a homogeneous astrometric dataset tied to a common physical reference, allowing a self-consistent fit of the coupled inner and outer orbits and direct constraints on the individual stellar masses. The resulting solutions establish the orbital framework needed for future dynamical studies of the circumtriple ring. Section~\ref{sec:observations} presents the new measurement. Sections~\ref{sec:oracle} and \ref{sec:Aa_conversion} describe the fitting framework and the construction of the $Aa$--$Ab_1$ astrometry. Section~\ref{sec:disk_constraint} introduces the ring-center constraint, and Sects.~\ref{sec:results} and \ref{sec:discussion} present and discuss the results.

\section{New $Aa$--$Ab_\mathrm{pc}$ astrometric measurement}
\label{sec:observations}

The orbital analysis presented in this work combines previously published astrometry with one new astrometric measurement of the wide GG~Tau~$A$ subsystem. The archival inner $Ab_2$--$Ab_1$ astrometry was adopted from \citet{Duchene2024}, while the archival wide $Aa$--$Ab_\mathrm{pc}$ astrometry was adopted from \citet{Toci2024}. The only new measurement considered here is an additional wide-orbit epoch obtained on 25 November 2024.

This new observation of the GG~Tau system was obtained with the Spectro-Polarimetric High-contrast Exoplanet REsearch instrument \citep[SPHERE;][]{Beuzit2019} at the Very Large Telescope (VLT), using the Infra-Red Dual-band Imaging and Spectroscopy instrument \citep[IRDIS;][]{Dohlen2008}, under program ID 114.27FG.001 (PI: C. Ginski). The dataset consists of eleven complete polarimetric cycles, each containing the four half-wave plate positions. Each polarimetric exposure has an integration time of 60 seconds (DIT$=$4, NDIT$=$15), resulting in a total on-source exposure time of 44 minutes. The data were reduced with the IRDIS Data reduction for Accurate Polarimetry pipeline \citep[IRDAP;][]{vanHolstein2020}, which provides an end-to-end reduction for SPHERE/IRDIS observations.

The relative astrometry of $Ab$ with respect to $Aa$ was extracted from the total-intensity image. The approximate locations of the two stellar components were first identified with a local peak-finding algorithm and then used as initial guesses for a simultaneous two-component 2D Gaussian fit performed with the Levenberg--Marquardt least-squares algorithm. The fitted Gaussian centroids provided the detector-frame positions of the two stars in pixel coordinates. The statistical uncertainties on the fitted parameters were estimated from the covariance matrix returned by the fit, whose diagonal elements correspond to the variances of the fitted parameters. The relative separation was computed from the differences in the fitted centroid positions and then converted into an on-sky value using the SPHERE/IRDIS $K$-band pixel scale ($12.265 \pm 0.009$ mas/pixel), yielding a projected separation of $258.67 \pm 1.08$ mas.

The position angle was derived from detector-frame centroid offsets using the appropriate two-argument arctangent function and then converted into degrees. We subsequently applied a true north correction of $-1.75^\circ$, the parallactic angle at the time of the flux observations, and the additional $135.99^\circ$ offset prescribed in the SPHERE manual \citep{SPHEREManualsESO}. This yields an on-sky position angle of $312.22 \pm 0.12^\circ$. The associated uncertainty was propagated from the partial derivatives of the position angle with respect to the fitted centroid positions and included a $0.1^\circ$ uncertainty on the true north correction. This new astrometric point extends the temporal coverage of the wide orbit, while the inner subsystem remains constrained by the previously published measurements alone.

\section{Orbital fitting with the \texttt{Oracle} code}
\label{sec:oracle}

The orbital analysis presented in this work was carried out with a dedicated fitting code, \texttt{Oracle}. Its statistical framework and sampling strategy rely on the Markov Chain Monte Carlo (MCMC) approach described by \citet{Ford2006} for orbital determination in extrasolar planetary systems. The purpose of \texttt{Oracle} is not to introduce a new fitting formalism but to extend this established framework to hierarchical stellar systems, in which multiple dynamically coupled orbits must be modeled simultaneously from heterogeneous observables.

\subsection{Scope of the \texttt{Oracle} code}
\label{subsec:oracle_scope}

\texttt{Oracle} is designed to fit astrometric data and, when available, radial-velocity data, within a unified Bayesian framework while simultaneously modeling multiple Keplerian orbits expressed in Jacobi coordinates. In hierarchical stellar systems such as GG~Tau~$A$, the orbital motions of the components are coupled through their masses. Treating the orbits independently therefore neglects part of the underlying dynamics and limits the physical information that can be extracted from the data.

Such coupling is generally negligible in planetary systems, where a dominant central mass hosts negligible-mass companions on nested orbits. In this regime, the barycenter of each inner subsystem lies very close to the central star, so the motion of an outer companion relative to that barycenter is nearly indistinguishable from its motion relative to the star itself. Independent orbital fits therefore usually provide an adequate approximation. In stellar multiple systems, by contrast, the component masses are comparable, and the barycenter of an inner subsystem can be substantially displaced from any individual star. Each outer orbit must consequently be defined relative to the barycenter of the corresponding inner subsystem, as required by the Jacobi-coordinate description \citep{Plummer1918,Beust2003}. The mutual gravitational influence can then produce measurable effects over the observational baseline.

In a joint multi-orbit fit, each orbit therefore provides a complementary constraint on the system masses. In GG~Tau~$A$, the wide orbit, defined by the motion of $Aa$ relative to $Ab_\mathrm{cm}$, constrains the total mass of the system, while the inner $Ab_1$--$Ab_2$ orbit constrains the total mass of the $Ab$ subsystem. The coupling between the two orbits further requires $Ab_\mathrm{cm}$ to follow the wide orbit self-consistently. This coupling introduces an additional constraint on the mass ratio between $Ab_1$ and $Ab_2$, making it possible to infer the individual stellar masses rather than only their sums.

Exploring multiple dynamically coupled orbital parameter spaces simultaneously nevertheless poses significant numerical challenges, especially when one of the orbits is only weakly constrained. In GG~Tau~$A$, the wide orbit remains poorly determined and admits solutions with long periods and high eccentricities. To sample such parameter spaces efficiently, \texttt{Oracle} implements a formulation of Keplerian motion based on universal variables \citep{DanbyBurkardt1983,BurkardtDanby1983,Danby1987}. This approach provides a continuous and numerically stable description of orbital motion across the full range of eccentricities \citep{Beust2016}. Compared with classical formulations based on true anomalies, universal variables improve sampling efficiency and ease transitions between low- and high-eccentricity solutions, as naturally encountered when a weakly constrained wide orbit is fitted jointly with a tighter inner binary.

\subsection{Model parameters and priors}
\label{subsec:oracle_parameters}

In the following, each Keplerian orbit is described by the standard set of orbital elements: the semimajor axis $a$, the eccentricity $e$, the inclination $i$, the argument of periastron $\omega$, the longitude of the ascending node $\Omega$, the time of periastron passage $t_p$, and the orbital period $P$. Although these elements provide the natural physical description of the orbit, they are not always the most convenient quantities for numerical sampling when the observational constraints are incomplete. The choice of fitted parameters and associated priors is therefore critical for the robustness and efficiency of orbital fitting.

With relative astrometry alone, orbital parameterization faces several well-known difficulties. The projected orbit remains invariant under the transformation $(\omega,\Omega)\rightarrow(\omega+\pi,\Omega+\pi)$, which means that astrometry by itself cannot distinguish the ascending node from the descending node. In addition, neither the semimajor axis nor the inclination is naturally suited to simple uniform sampling, and angular parameters more generally can introduce artificial discontinuities at the boundaries of their definition intervals.

For the weakly constrained wide orbit considered here, the universal-variable formulation is particularly useful because the posterior may extend toward highly eccentric or even formally unbound solutions. In this framework, the semimajor axis $a$ is replaced by the periastron distance $q$, which remains well defined for both bound and unbound trajectories. The characteristic timescale is described by $P_q$, defined as the period of a circular orbit of radius $q$, while the orbital phase is specified by the universal variable $s$, which provides a continuous description of the position along the orbit.

Accordingly, and consistently with the parameterization adopted by \citet{Beust2016}, the fitted orbital parameters are $\ln(q)$, $e\cos(\omega+\Omega)$, $e\sin(\omega+\Omega)$, $\sin(i/2)\cos(\omega-\Omega)$, $\sin(i/2)\sin(\omega-\Omega)$, $\ln(P_q)$, and $s$. Unless otherwise stated, uniform priors are adopted on all fitted parameters in their transformed space. This choice is intended to limit parameter degeneracies, ensure a smooth exploration of parameter space, and provide a robust description of poorly constrained orbits.

In addition to orbital parameters, \texttt{Oracle} allows the inclusion of external priors on combinations of stellar masses when independent observational constraints are available. Such priors may come from independent stellar-mass estimates or from dynamical measurements external to the orbital fit itself. In the specific case of GG~Tau~$A$, the total mass of the triple system is independently constrained by measurements of the Keplerian rotation of the circumtriple disk. We incorporated this information through a Gaussian prior on the total system mass, $M_{\mathrm{tot}} = 1.41 \pm 0.08\,M_\odot$, obtained by rescaling the mass derived by \citet{Phuong2020} to the distance adopted in this work. This prior was applied consistently within the joint multi-orbit fit, and its uncertainty was naturally propagated into the inferred orbital parameters and individual stellar masses.

The likelihood function adopted in \texttt{Oracle} follows the Bayesian formalism described by \citet{Ford2005}. Under the assumption that the model provides the exact underlying signal and that the measurement errors are Gaussian, the fit is based on the usual $\chi^2$ statistic, here denoted $\chi^2_{\rm orb}$. For relative astrometric measurements, this $\chi^2_{\rm orb}$ is evaluated in 2D on the plane of the sky. When available, correlations between astrometric observables are explicitly accounted for through the full covariance matrices associated with the measurements; otherwise, the uncertainties are assumed to be uncorrelated. This formulation provides a consistent and flexible framework for combining heterogeneous astrometric datasets within the orbital fitting procedure.

\section{Conversion of $Aa$--$Ab_\mathrm{pc}$ into $Aa$--$Ab_1$ astrometry}
\label{sec:Aa_conversion}

Historical astrometric observations of the wide orbit of GG~Tau~$A$ do not resolve the individual components of the $Ab_1$--$Ab_2$ subsystem but instead measure the position of its center of light, $Ab_\mathrm{pc}$. These measurements therefore provide relative astrometry between the $Aa$ component and a time-dependent observational reference, rather than between two physical stellar components.

Modeling GG~Tau~$A$ as a hierarchical triple requires all astrometric measurements to be expressed in a common and physically meaningful reference frame. In particular, a self-consistent joint fit of the inner and wide orbits requires all relative positions to be referred to the same point. Although the photocenter $Ab_\mathrm{pc}$ provides a convenient observational reference for unresolved measurements, it does not correspond to a physical object with a well-defined equation of motion. By contrast, the inner orbit is naturally described by the position of $Ab_2$ relative to $Ab_1$, which makes $Ab_1$ a physically meaningful reference shared by both the inner and wide orbital architectures. For this reason, all wide-orbit astrometric measurements are converted below to be referenced to $Ab_1$. This observational choice should not be confused with the dynamical parameterization of the hierarchical model itself. The wide orbit is still defined by the motion of $Aa$ relative to $Ab_\mathrm{cm}$. Expressing the astrometry as $Aa$--$Ab_1$ measurements simply reflects a common reference frame for the joint fitting of the inner and wide orbits, as is standard practice in the framework of Jacobi coordinates.

\subsection{$Ab$ photocenter definition and reconstruction}
\label{subsec:Aa_conversion_photocenter}

For an unresolved binary such as the $Ab_1$--$Ab_2$ pair, the photocenter corresponds to the flux-weighted center of light of the system. In general, it coincides with neither stellar component nor, a priori, the center of mass. Its position depends on both the instantaneous orbital configuration of the binary and on the flux ratio between its components, and therefore varies with time and wavelength.

In the specific case of GG~Tau~$A$, this distinction is particularly important. The photocenter of the $Ab_1$--$Ab_2$ subsystem undergoes a time-variable displacement driven by the orbital motion of the inner binary. If this effect were neglected, part of the photocenter motion induced by the inner binary would be incorrectly absorbed into the wide-orbit solution, leading to systematic biases in the inferred wide-orbit parameters.

Rather than defining the photocenter through absolute positions of the individual components, it is more natural in the present context to describe it directly as an offset relative to the physical component $Ab_1$. Denoting by $F_1$ and $F_2$ the fluxes of $Ab_1$ and $Ab_2$ in a given spectral band, respectively, the photocenter offset relative to $Ab_1$ can be written as
\begin{equation}
\overrightarrow{Ab_1\,Ab_\mathrm{pc}}
=
\frac{F_2/F_1}{1 + F_2/F_1}\,
\overrightarrow{Ab_1\,Ab_2}.
\label{eq:photocenter_vector}
\end{equation}
This formulation is equivalent to the standard flux-weighted definition of the photocenter used for unresolved binaries in astrometry and describes the displacement that unresolved companions induce in modern space-based astrometric measurements \citep{Belokurov2020,Halbwachs2023}.

For the GG~Tau~$Ab$ subsystem, \citet{Duchene2024} report four independent measurements of the $K$-band flux ratio between $Ab_2$ and $Ab_1$. Although young stars can exhibit photometric variability, it is generally less pronounced in the $K$ band than at optical wavelengths \citep{KenyonHartmann1995}. In the specific case of GG~Tau~$Ab$, the four measured flux ratios are mutually consistent within their uncertainties, indicating little if any variability in this system, either over time or across the $K$ band. We therefore adopted a single representative value given by the inverse-variance weighted mean of these measurements:
\begin{equation}
\left(\frac{F_2}{F_1}\right)_K = 0.19 \pm 0.02 .
\end{equation}

In contrast, the flux ratio remains poorly constrained in other spectral bands used by some historical astrometric measurements. To avoid introducing additional poorly constrained assumptions, the analysis presented here is therefore restricted to $Aa$--$Ab_\mathrm{pc}$ measurements obtained in the $K$ band.

\subsection{Intermediate $Ab_1$--$Ab_2$ binary solution}
\label{subsec:Aa_conversion_inner_solution}

Converting the wide-orbit astrometric measurements from an $Aa$--$Ab_\mathrm{pc}$ reference to an $Aa$--$Ab_1$ reference requires an estimate of the relative orbital motion of the inner $Ab_1$--$Ab_2$ binary. This is because the photocenter displacement at a given epoch depends on the instantaneous position of $Ab_2$ with respect to $Ab_1$.

To obtain this information, we derived an intermediate orbital solution for the $Ab_1$--$Ab_2$ system using the \texttt{Oracle} fitting framework described in Sect.~\ref{sec:oracle}. This fit relies only on the astrometric measurements of the inner binary and was performed independently of the wide-orbit data. At each observation epoch, this fit yields a posterior distribution for the relative position of $Ab_2$ with respect to $Ab_1$, thereby allowing the uncertainties and parameter correlations to be accounted for. A representative relative position was then derived for each epoch from the median of this distribution, together with an associated uncertainty given by its standard deviation.

This procedure is justified because the inner $Ab_1$--$Ab_2$ orbit is already well constrained by the available astrometric data. The existing observations span a full orbital cycle of the inner binary, providing a precise determination of its orbital period and strongly restricting the range of admissible orbital solutions. As a result, the uncertainty on the reconstructed photocenter displacement remains small compared with the measurement uncertainties of the wide-orbit astrometry. The resulting intermediate solution also agrees within the uncertainties with the orbital solution published by \citet{Duchene2024}.

This intermediate solution was used solely to reconstruct the photocenter motion of the $Ab_1$--$Ab_2$ subsystem and express the wide-orbit astrometry in a reference frame tied to $Ab_1$, as described in the following subsection. It was not intended as a final characterization of the inner binary, whose orbital parameters and stellar masses were derived from the joint fit of the inner and wide orbits presented in Sect.~\ref{sec:results}.

\subsection{Construction of the $Aa$--$Ab_1$ astrometry}
\label{subsec:Aa_conversion_construction}

Using the notation introduced above, the relative position of $Aa$ with respect to $Ab_1$ at a given epoch was obtained by combining the observed astrometry of $Aa$ relative to the photocenter $Ab_\mathrm{pc}$ with the offset of the photocenter relative to $Ab_1$. The measured $Aa$--$Ab_\mathrm{pc}$ astrometry can thus be converted into $Aa$--$Ab_1$ astrometry by accounting for the instantaneous photocenter displacement induced by the inner binary.

The uncertainties on the two terms entering this expression were propagated under the assumption that the observed $Aa$--$Ab_\mathrm{pc}$ astrometry and the reconstructed photocenter offset relative to $Ab_1$ were statistically independent. Under this assumption, the two contributions were added in quadrature. In practice, the uncertainty budget is dominated by the measurement errors of the historical relative astrometry, which justifies the reconstruction procedure adopted here.

\begin{figure}[htbp]
\centering
\includegraphics[width=0.85\columnwidth]{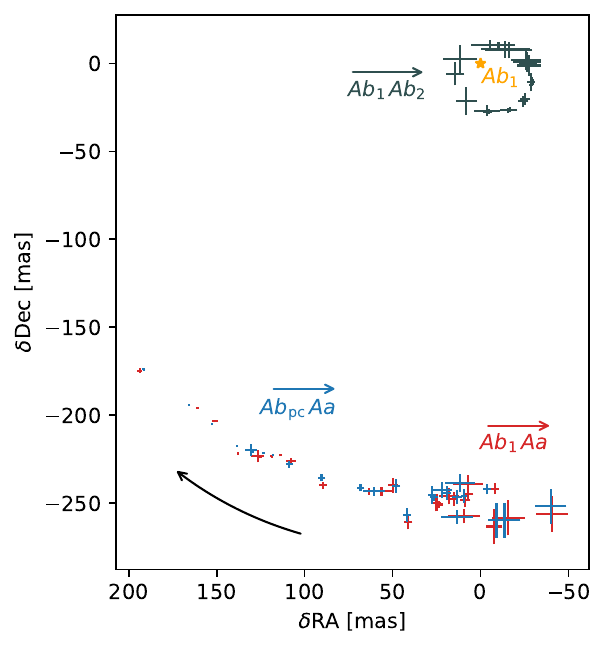}
\caption{Comparison between the original $Aa$--$Ab_\mathrm{pc}$ astrometry and the reconstructed $Aa$--$Ab_1$ astrometry. The blue points show the observed positions of $Aa$ relative to $Ab_\mathrm{pc}$, while the orange star marks $Ab_1$, adopted as the origin of the reference frame. At the same epochs, the reconstructed positions of $Ab_2$ relative to $Ab_1$ are shown in gray, and the corresponding positions of $Aa$ relative to $Ab_1$, obtained by accounting for the photocenter offset, are shown in red. Each red point is therefore directly associated with one observed blue point at the same epoch. The black arrow indicates the direction of motion along the orbit and hence the direction of increasing time. The offset between the blue and red positions is small compared with the uncertainties of the oldest measurements. However, it becomes significant for the most recent high-precision epochs.}
\label{fig:aa_abpc_vs_ab1}
\end{figure}

Figure~\ref{fig:aa_abpc_vs_ab1} compares the original $Aa$--$Ab_\mathrm{pc}$ astrometry with the reconstructed $Aa$--$Ab_1$ measurements. The offset, which ranges from 2 to 5 mas, is small compared with the uncertainties of the oldest epochs but becomes significant for the most recent high-precision measurements. The resulting homogeneous set of $Aa$--$Ab_1$ relative astrometric measurements forms the basis of the wide-orbit analysis presented in Table~\ref{tab:aa_ab1_astrometry}.

\section{Constraint from the circumtriple disk}
\label{sec:disk_constraint}

The circumtriple disk provides three complementary observational constraints on the central triple system. The first, derived from its Keplerian rotation curve, is already included in the orbital fit through a prior on the total stellar mass, $M_{\mathrm{tot}} = M_{Aa} + M_{Ab_1} + M_{Ab_2}$, where $M_{Aa}$, $M_{Ab_1}$, and $M_{Ab_2}$ denote the masses of $Aa$, $Ab_1$, and $Ab_2$, respectively, introduced in Sect.~\ref{subsec:oracle_parameters}. A second constraint is provided by the geometrical center of the ring in the plane of the sky, and a third is provided by the observed inner edge of the disk, or equivalently by the cavity size, although this last constraint is left aside in the present paper.

Although these three constraints arise from the same physical structure, they probe different aspects of it: the rotation curve constrains the gravitational potential through kinematics; the ring center provides a purely geometrical measurement; and the inner edge probes the dynamical response of the disk to the stellar architecture. Using these observables is therefore not redundant but instead provides complementary tests of the consistency between the stellar orbital solution and the large-scale properties of the disk. In the following, we focus on the second of these observables, namely the geometrical center of the ring, whose constraint is applied in post-processing to the posterior of the orbital fit presented in Sect.~\ref{sec:results}.

\subsection{Physical basis of the ring-center constraint}

The observed velocity field of the circumtriple disk is well described by Keplerian rotation around a single dynamical center \citep{Guilloteau1999,Phuong2018}, indicating that the disk responds primarily to the large-scale gravitational potential of the central triple. In addition, once deprojected into the disk plane, the disk appears approximately axisymmetric and hence nearly circular on large scales, despite the presence of localized substructures within the cavity and along the disk \citep{Dutrey1994,Guilloteau1999,Dutrey2014,Phuong2018,Phuong2020}. Under this working assumption, the geometrical center derived from ring modeling is expected to coincide with the projected barycenter of the central triple. We therefore did not treat the measured ring center as an exact dynamical observable, unlike the total mass of the triple. Instead, we regard it as an independent geometrical constraint, whose interpretation may be affected by residual asymmetries in the disk structure.

The ring center measured by \citet{Phuong2018} from Atacama Large Millimeter/submillimeter Array (ALMA) observations of the dust continuum at $0.9$~mm, obtained on 18 November 2013, is denoted $G_{\mathrm{ring}}$. Its position is given relative to the compact continuum emission associated with $Aa$ as
\begin{equation}
\overrightarrow{Aa\,G_{\mathrm{ring}}}
=
\left(
\delta \mathrm{RA}_{\mathrm{ring}},
\delta \mathrm{Dec}_{\mathrm{ring}}
\right),
\end{equation}
with $\delta \mathrm{RA}_{\mathrm{ring}} = -110 \pm 30$~mas and $\delta \mathrm{Dec}_{\mathrm{ring}} = 40 \pm 30$~mas. This compact continuum emission traces circumstellar dust around $Aa$ rather than the stellar photosphere directly. In the following, we assume that its centroid coincides with the position of $Aa$ and neglect any additional systematic uncertainty associated with a possible offset between the star and its circumstellar dust emission. The uncertainties on the two coordinates are assumed to be independent and Gaussian.

\subsection{Computation of the triple center of mass}

In the following, all projected positions are expressed relative to $Aa$ in order to remain strictly consistent with the ring-center measurement and avoid introducing additional uncertainty through coordinate transformations. This local choice of reference is specific to the ring-center constraint and does not affect the earlier use of $Ab_1$ for constructing the wide-orbit astrometry in Sect.~\ref{sec:Aa_conversion}, since the two steps relied on different observables.

For each posterior orbital solution derived in Sect.~\ref{subsec:fit_results_without_ring}, the projected positions of $Aa$, $Ab_1$, and $Ab_2$ were evaluated at the epoch of the ring-center measurement. The projected barycentric offset $A_\mathrm{cm}$ of the system with respect to $Aa$ was then given by
\begin{equation}
\overrightarrow{Aa\,A_\mathrm{cm}}
=
\frac{
M_{Ab_1}\,\overrightarrow{Aa\,Ab_1}
+
M_{Ab_2}\,\overrightarrow{Aa\,Ab_2}
}{
M_{Aa} + M_{Ab_1} + M_{Ab_2}
}.
\end{equation}
Each posterior sample therefore predicted a center-of-mass position in the plane of the sky.

To quantify the consistency of the predicted barycentric position with the observed ring-center position, we defined for each sample $i$ a $\chi^2_{{\rm cm},i}$ quantity. This quantity was the quadratic sum, over the two sky-plane coordinates, of the residuals between the predicted barycentric position and the observed ring-center position, weighted by the corresponding observational uncertainties. Rather than incorporating this observable directly into the orbital fit, we used it as a secondary a posteriori constraint in post-processing by assigning each sample a weight $\exp\left(-\chi^2_{{\rm cm},i}/2\right)$. This was equivalent to multiplying the sampled posterior by the Gaussian likelihood associated with the ring-center measurement, while leaving the original exploration of parameter space unchanged. 

This procedure is approximate, as it assumes that the ring-center measurement is statistically independent of the stellar astrometry and that its quoted uncertainties adequately capture the uncertainty on the geometrical-center estimate. It nevertheless provides a simple way to assess which orbital solutions remain compatible with the large-scale disk geometry. Its impact on the posterior distribution and on the inferred orbital solution is presented in Sect.~\ref{subsec:ring_constraint_results}.

\section{Results: Orbital architecture of GG~Tau~$A$}
\label{sec:results}

We first consider the hierarchical triple fit including the new wide-orbit astrometric measurement (Case~A), before examining the effect of the additional ring-center constraint applied in post-processing (Case~B). Figures~\ref{fig:corner_inner} and \ref{fig:corner_outer} show the posterior distributions of the orbital parameters for both cases for the inner and outer orbits, respectively. These corner plots provide an immediate visual assessment of how the ring-center constraint sharpens the posterior distributions and more tightly constrains the system, while preserving both the marginal distributions and the correlations between parameters. Table~\ref{tab:orbital_parameters}, by contrast, provides a compact summary in terms of posterior medians and credible intervals for the orbital parameters.

{\renewcommand{\arraystretch}{1.5}
\begin{table}[htbp]
\setlength{\tabcolsep}{4pt}
\centering
\caption{Summary of the inferred orbital parameters.}
\label{tab:orbital_parameters}
\begin{tabular*}{\columnwidth}{@{\extracolsep{\fill}}l c r@{\hspace{-1.7em}}l r@{\hspace{-1.7em}}l@{}}
\hline\hline
Parameter & Unit & \multicolumn{2}{c}{Case~A} & \multicolumn{2}{c}{Case~B} \\
\hline
$P_{\rm in}$ & yr & $8.055$ & ${}^{+0.093}_{-0.095}$ & $8.056$ & ${}^{+0.094}_{-0.093}$ \\
$a_{\rm in}$ & au & $3.53$ & ${}^{+0.25}_{-0.16}$ & $3.44$ & ${}^{+0.15}_{-0.11}$ \\
$e_{\rm in}$ & -- & $0.459$ & ${}^{+0.049}_{-0.032}$ & $0.450$ & ${}^{+0.035}_{-0.027}$ \\
$i_{\rm in}$ & deg & $149$ & ${}^{+11}_{-10}$ & $154.2$ & ${}^{+10.0}_{-8.6}$ \\
$\omega_{\rm in}$ & deg & $244$ & ${}^{+29}_{-16}$ & $241$ & ${}^{+31}_{-18}$ \\
$\Omega_{\rm in}$ & deg & $94$ & ${}^{+45}_{-21}$ & $95$ & ${}^{+41}_{-23}$ \\
$t_{p,\rm in}$ & MJD & $57945$ & ${}^{+178}_{-110}$ & $57939$ & ${}^{+108}_{-68}$ \\
\hline
$P_{\rm out}$ & yr & $237$ & ${}^{+136}_{-51}$ & $239$ & ${}^{+140}_{-53}$ \\
$a_{\rm out}$ & au & $42.9$ & ${}^{+15.3}_{-6.4}$ & $43.2$ & ${}^{+15.7}_{-6.6}$ \\
$e_{\rm out}$ & -- & $0.230$ & ${}^{+0.147}_{-0.054}$ & $0.235$ & ${}^{+0.150}_{-0.053}$ \\
$i_{\rm out}$ & deg & $138.5$ & ${}^{+6.2}_{-6.0}$ & $138.0$ & ${}^{+6.1}_{-5.9}$ \\
$\omega_{\rm out}$ & deg & $48$ & ${}^{+44}_{-31}$ & $48$ & ${}^{+43}_{-30}$ \\
$\Omega_{\rm out}$ & deg & $116.7$ & ${}^{+8.1}_{-11.9}$ & $116.7$ & ${}^{+7.9}_{-11.5}$ \\
$t_{p,\rm out}$ & MJD & $72202$ & ${}^{+7900}_{-6900}$ & $72039$ & ${}^{+7700}_{-6700}$ \\
\hline
\end{tabular*}
\tablefoot{The values listed here are given as posterior medians with 16th--84th percentile intervals. The orbital elements are expressed in Jacobi coordinates: inner-orbit parameters refer to $Ab_2$ relative to $Ab_1$, and outer-orbit parameters to $Aa$ relative to $Ab_\mathrm{cm}$. Only one of the two equivalent branches of the astrometric degeneracy is shown. The complementary solution $(\omega+\pi, \Omega+\pi)$ produces the same projected orbit and is therefore equally valid.}
\end{table}
}

\subsection{Triple fit including the new wide-orbit astrometric epoch}
\label{subsec:fit_results_without_ring}

Case~A relies on the full methodology introduced earlier. The historical wide-orbit astrometry, together with the new wide-orbit measurement, was converted from $Aa$--$Ab_\mathrm{pc}$ to $Aa$--$Ab_1$ measurements following Sect.~\ref{sec:Aa_conversion}. The resulting dataset was then fitted jointly with the inner $Ab_2$--$Ab_1$ astrometry within the hierarchical \texttt{Oracle} framework described in Sect.~\ref{sec:oracle}. This simultaneous treatment consistently propagated the coupling between the two orbits through the mass budget of the system.

Because our analysis simultaneously fits the hierarchical architecture whereas \citet{Duchene2024} and \citet{Toci2024} fitted the inner and wide orbits separately, respectively, a direct one-to-one comparison of the posterior estimates should be interpreted with caution, especially for the wide orbit. For the inner binary, Case~A yields $P_{\rm in}=8.055^{+0.093}_{-0.095}$~yr and $a_{\rm in}=3.53^{+0.25}_{-0.16}$~au, close to the values of $P_{\rm in}=8.16^{+0.08}_{-0.07}$~yr and $a_{\rm in}=3.7^{+0.1}_{-0.1}$~au obtained by \citet{Duchene2024}. The inferred eccentricity, $e_{\rm in}=0.459^{+0.049}_{-0.032}$, is lower than their value of $e_{\rm in}=0.539^{+0.033}_{-0.027}$. For the wide orbit, Case~A gives $P_{\rm out}=237^{+136}_{-51}$~yr, $a_{\rm out}=42.9^{+15.3}_{-6.4}$~au, and $e_{\rm out}=0.230^{+0.147}_{-0.054}$, broadly similar to the posterior estimates of \citet{Toci2024}, namely $P_{\rm out}=199^{+67}_{-28}$~yr, $a_{\rm out}=38^{+8}_{-3}$~au, and $e_{\rm out}=0.22^{+0.06}_{-0.05}$. Given the still incomplete coverage of the wide orbit, the additional astrometric epoch only modestly affects the inferred orbital architecture.

\subsection{Refinement from the ring-center constraint}
\label{subsec:ring_constraint_results}

Case~B was obtained by applying the ring-center constraint to the Case~A posterior through the reweighting scheme described in Sect.~\ref{sec:disk_constraint}. This post-processing step leaves the original Case~A results unchanged, while allowing the specific impact of the ring-center information on the posterior distributions to be assessed directly.

\begin{figure}[htbp]
    \centering
    \includegraphics[width=0.85\columnwidth]{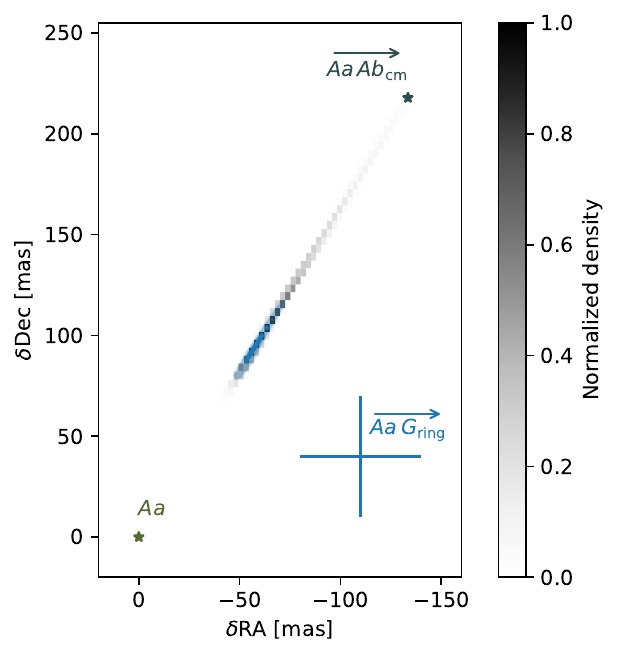}
    \caption{Effect of the constraint from the geometrical center of the ring on the posterior distribution of the projected position of the triple-system barycenter in the plane of the sky. The posterior distribution is shown in gray before reweighting and as blue contours after application of this constraint. The contours enclose 12\%, 39\%, 68\%, and 87\% of the total posterior probability, with increasing opacity toward regions of higher posterior density. The blue cross marks the measured position of the ring center relative to $Aa$. The reweighted distribution is therefore more concentrated around this position.}
    \label{fig:ring_constraint}
\end{figure}

Figure~\ref{fig:ring_constraint} shows that the reweighting narrows the posterior distribution of the projected center-of-mass position in the plane of the sky, as expected, and shifts its highest-probability region toward the measured ring-center position. The ring-center information therefore acts mainly as a geometrical filter, selecting barycentric configurations consistent with the observed large-scale disk geometry without substantially changing the preferred orbital architecture. In particular, the posterior distributions of the wide-orbit parameters remain nearly unchanged. Although the ring-center constraint slightly sharpens the posterior distributions of the inner-orbit parameters, it leaves their medians essentially unchanged, yielding $P_{\rm in}=8.056^{+0.094}_{-0.093}$~yr, $a_{\rm in}=3.44^{+0.15}_{-0.11}$~au, and $e_{\rm in}=0.450^{+0.035}_{-0.027}$ in Case~B. Its main contribution is instead to restrict the allowed barycentric configurations and thus the mass partition within the triple. In particular, configurations in which the barycenter of the triple lies very close to the center of mass of the $Ab$ subsystem are disfavored, thereby ruling out solutions with very small values of the mass of $Aa$, $M_{Aa}$.

\begin{figure*}[htbp]
\centering
\sidecaption
\includegraphics[width=12cm]{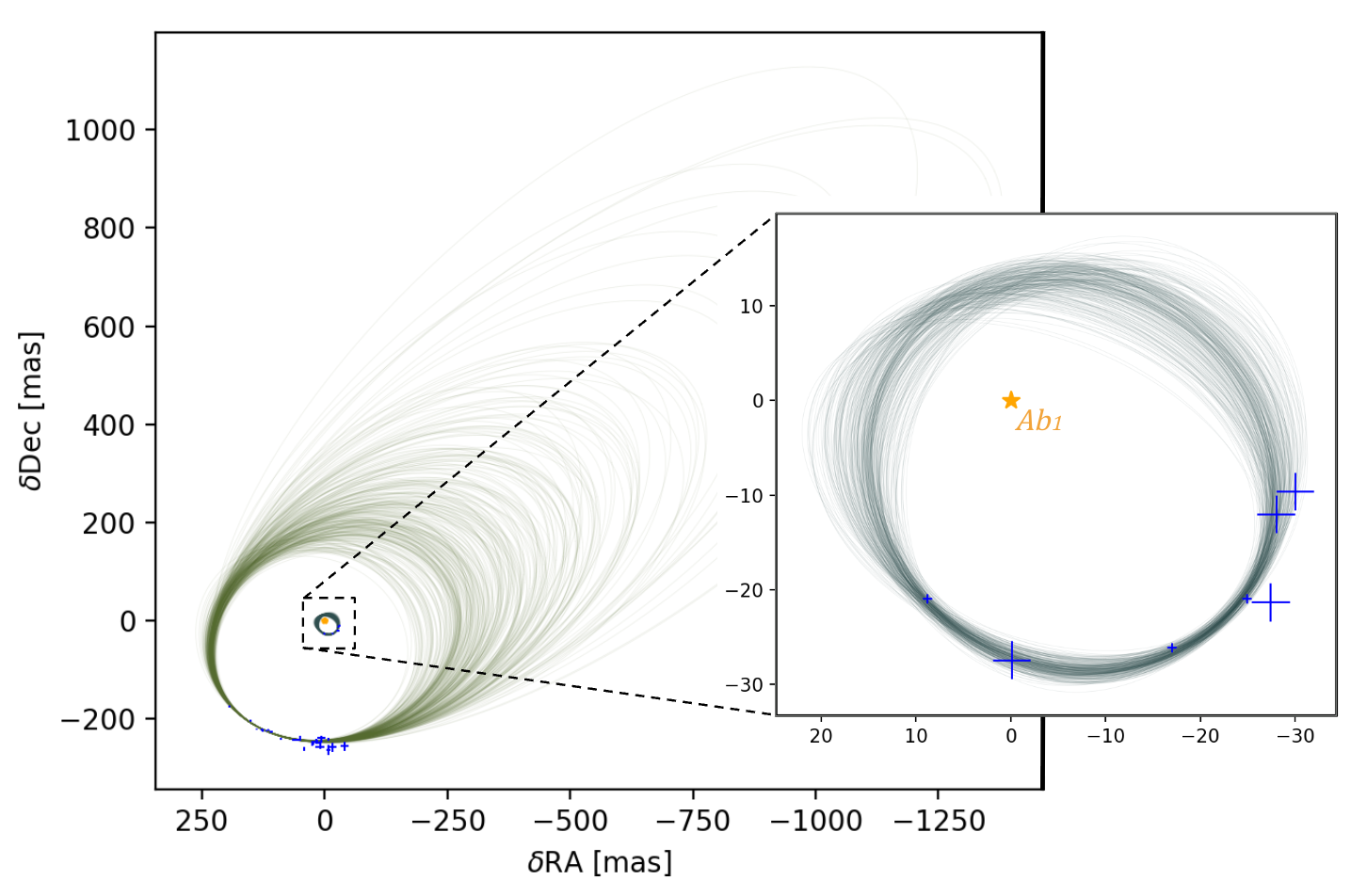}
\caption{Projected orbital architecture of GG~Tau~$A$ in the plane of the sky for Case~A, based on a composite visualization of 300 randomly selected posterior samples. Case~B is very similar and is therefore not shown. The inner orbit, shown in gray, traces the motion of $Ab_2$ relative to $Ab_1$, whereas the outer orbit, shown in green, traces that of $Aa$ relative to the center of mass of the $Ab_1$--$Ab_2$ subsystem. The blue crosses indicate the astrometric measurements.}
\label{fig:architecture_3d}
\end{figure*}

This final solution provides the first orbital reconstruction of GG~Tau~$A$ that is simultaneously consistent with the inner astrometry, the wide astrometry, and the geometrical and kinematic constraints derived from the circumtriple disk. A projected view in the plane of the sky of 300 randomly selected posterior samples is shown in Fig.~\ref{fig:architecture_3d}.

\subsection{Individual masses}

\begin{figure*}[htbp]
\centering
\sidecaption
\begin{minipage}{12cm}
  \centering
  \includegraphics[width=12cm]{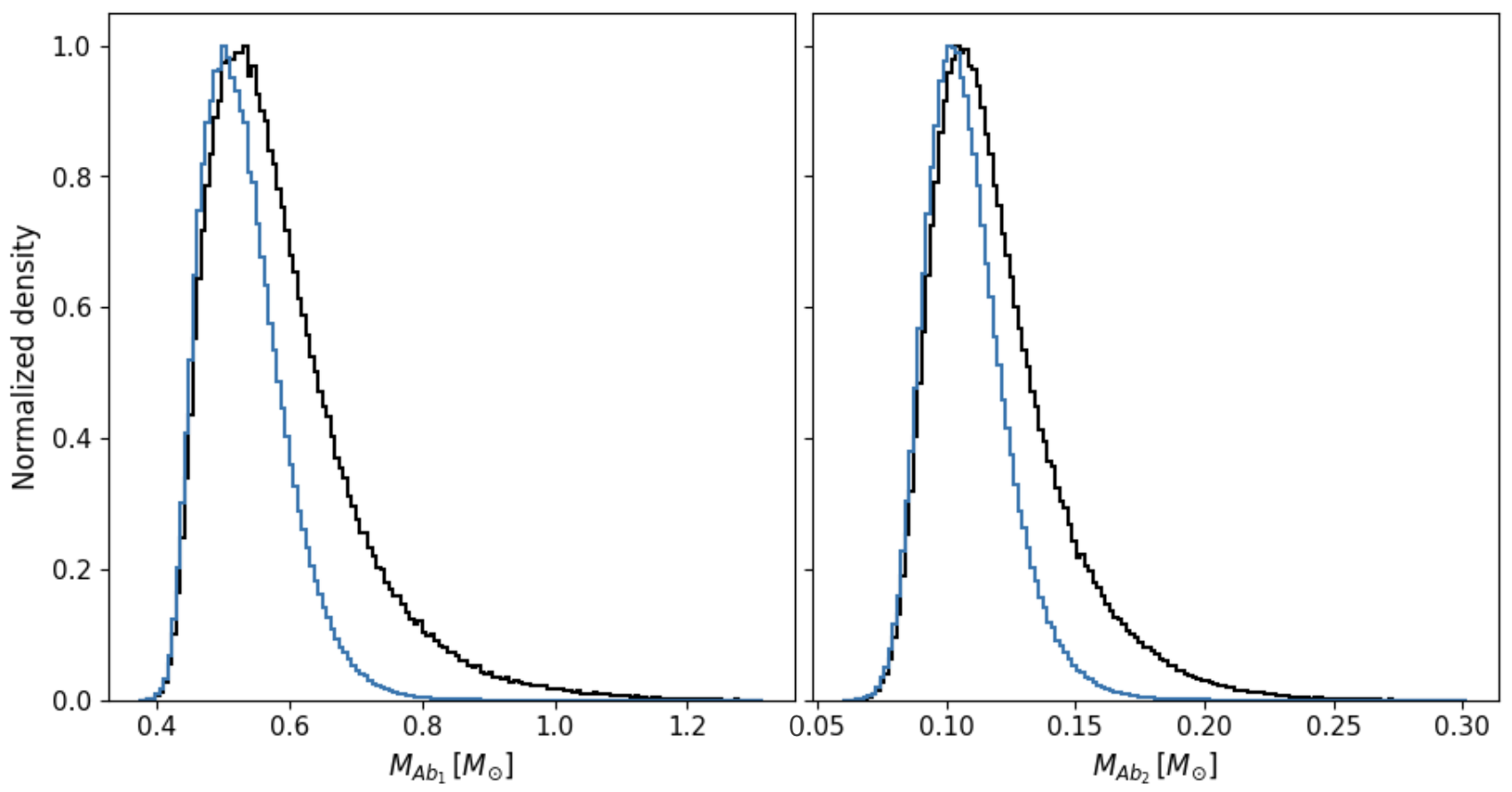}
  \vspace{0.5em}
  \includegraphics[width=12cm]{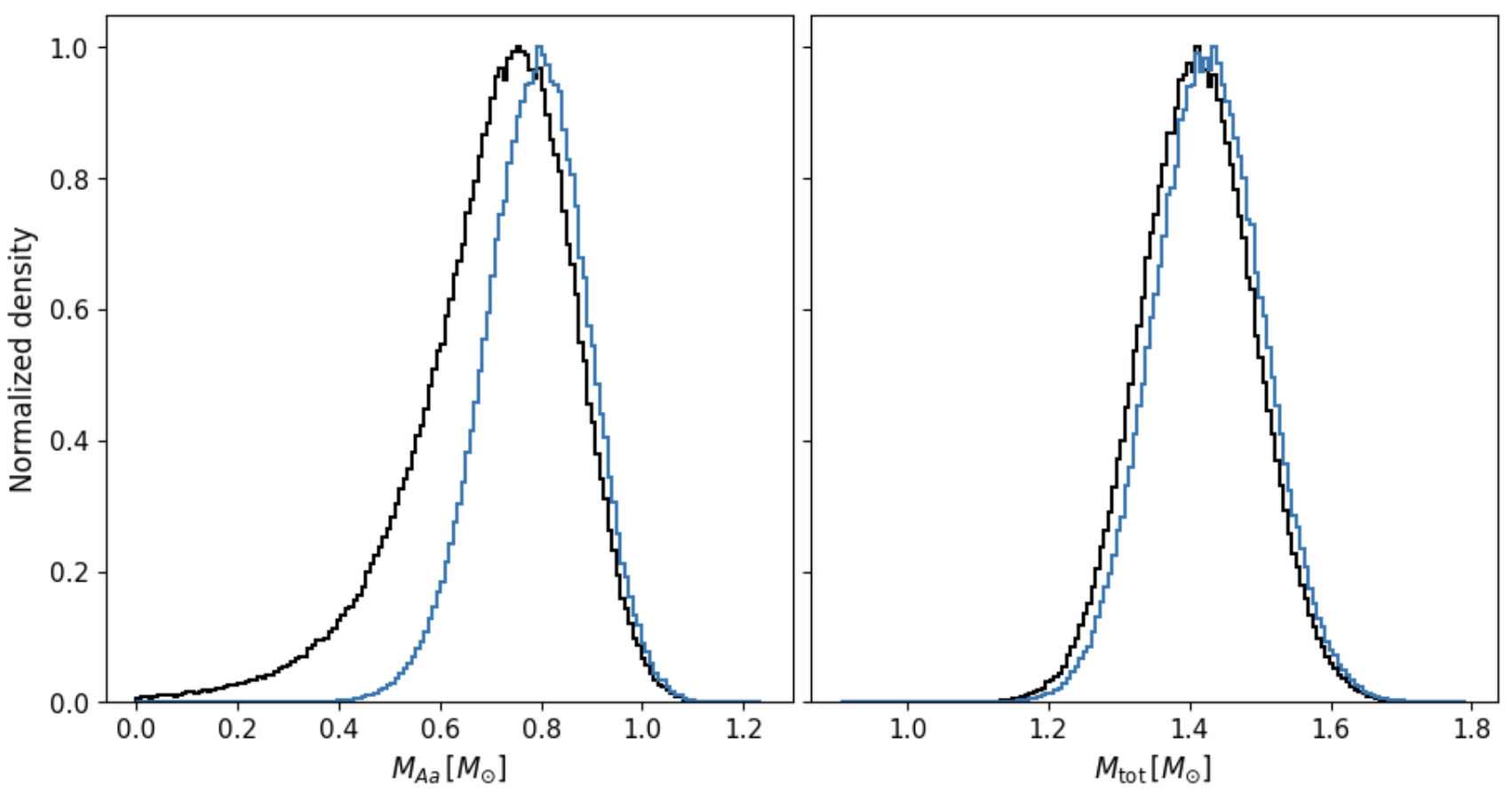}
\end{minipage}
\caption{Posterior distributions of the stellar masses for Cases~A and B, shown in black and blue, respectively. From top to bottom and left to right: Distributions of $M_{Ab_1}$, $M_{Ab_2}$, $M_{Aa}$, and $M_{\mathrm{tot}}$. The panels illustrate how the constraint from the ring center further tightens the mass constraints derived from the hierarchical triple fit.}
\label{fig:mass_posteriors}
\end{figure*}

The main scientific gain of the full analysis is the inference of the individual stellar masses, whose posterior distributions are shown in Fig.~\ref{fig:mass_posteriors}. This inference would not be possible without the simultaneous triple fit. Because the observations considered here are purely relative astrometric measurements, they do not directly constrain how the total mass is partitioned within the system. This information emerges only through the simultaneous modeling of the inner and wide orbits, which couples the two orbital levels through a common mass budget.

The corresponding values are listed in Table~\ref{tab:mass_budget}. Case~A already constrains the mass partition and provides estimates of the individual stellar masses $M_{Aa}$, $M_{Ab_1}$, and $M_{Ab_2}$. Case~B, which further incorporates the ring-center constraint, tightens these constraints by restricting the barycentric configurations compatible with the observed ring geometry. Consequently, the posterior distributions of the individual masses and their ratios become measurably narrower. As expected, however, this additional constraint leaves the total mass $M_{\mathrm{tot}}$ essentially unchanged, since it is already tightly constrained by the orbital fit and the adopted dynamical prior and because the barycentric position depends primarily on how the mass is distributed among the three components and only weakly on the total mass itself.

{\renewcommand{\arraystretch}{1.4}
\begin{table}[htbp]
\setlength{\tabcolsep}{4pt}
\centering
\caption{Summary of the inferred stellar masses and mass ratios.}
\label{tab:mass_budget}
\begin{tabular*}{\columnwidth}{@{\extracolsep{\fill}}l c r@{\hspace{-1.7em}}l r@{\hspace{-1.7em}}l@{}}
\hline\hline
Mass & Unit & \multicolumn{2}{c}{Case~A} & \multicolumn{2}{c}{Case~B} \\
\hline
$M_{\rm tot}$ & $M_\odot$ & $1.412$ & ${}^{+0.080}_{-0.079}$ & $1.427$ & ${}^{+0.078}_{-0.077}$ \\
$M_{Aa}$ & $M_\odot$ & $0.73$ & ${}^{+0.12}_{-0.17}$ & $0.79$ & ${}^{+0.10}_{-0.10}$ \\
$M_{Ab_1}$ & $M_\odot$ & $0.563$ & ${}^{+0.123}_{-0.075}$ & $0.521$ & ${}^{+0.069}_{-0.051}$ \\
$M_{Ab_2}$ & $M_\odot$ & $0.115$ & ${}^{+0.029}_{-0.017}$ & $0.106$ & ${}^{+0.017}_{-0.013}$ \\
$M_{Ab_2}/M_{Ab_1}$ & -- & $0.204$ & ${}^{+0.022}_{-0.021}$ & $0.203$ & ${}^{+0.021}_{-0.020}$ \\
$M_{Aa}/M_{Ab_1}$ & -- & $1.29$ & ${}^{+0.41}_{-0.47}$ & $1.52$ & ${}^{+0.32}_{-0.33}$ \\
\hline
\end{tabular*}
\tablefoot{The values listed here are given as posterior medians with 16th--84th percentile intervals. Mass ratios were computed directly from the posterior samples and therefore retain the correlations between the individual stellar masses.}
\end{table}
}

These mass estimates also remain qualitatively consistent with the spectroscopic analysis of \citet{White1999} and \citet{Hartigan2003}. In particular, if the composite spectrum of the $Ab$ subsystem is dominated by $Ab_1$, as suggested by the observed flux contrast, then the inferred mass ratio between $Aa$ and $Ab_1$ is in good agreement with the corresponding spectroscopic estimate. This agreement therefore provides an independent consistency check on the mass partition derived from the hierarchical orbital fit.

An interesting result, already visible in Case~A, is that the inferred mass ratio between $Ab_2$ and $Ab_1$ is very close to the $K$-band flux ratio adopted to reconstruct the photocenter motion in Sect.~\ref{subsec:Aa_conversion_photocenter}. The photocenter of the inner pair is therefore expected to lie close to its center of mass. In the specific case of GG~Tau~$Ab$, this shows that using the photocenter of the inner pair as an approximation to its barycenter in wide-orbit astrometry, as in \citet{Toci2024}, is justified a posteriori, even though these two quantities are not expected to coincide a priori.

\subsection{Mutual orientations and dynamical families}
\label{subsec:mutual_inclinations}

Beyond the orbital elements and masses, the relative orientations of the angular-momentum vectors provide additional information on the architecture of the system. We therefore examined the posterior distributions of two derived angles: the mutual inclination $\mathcal{I}_{\mathrm{in-out}}$ between the inner and wide orbits, and the relative inclination $\mathcal{I}_{\mathrm{triple-ring}}$ between the total angular momentum of the triple and that of the circumtriple ring. 

For the circumtriple ring of GG~Tau~$A$, we adopted an inclination of $i_{\rm ring}=35^\circ$ and a longitude of the ascending node $\Omega_{\rm ring}=97^\circ$, following the geometry derived from millimeter interferometric observations of the disk \citep{Guilloteau1999,Phuong2020}. In our orbital convention, this inclination is written as $180^\circ-i_{\rm ring}=145^\circ$ to select the corresponding orientation of the angular momentum vector.

\begin{figure*}[htbp]
\centering
\sidecaption
\includegraphics[width=12cm]{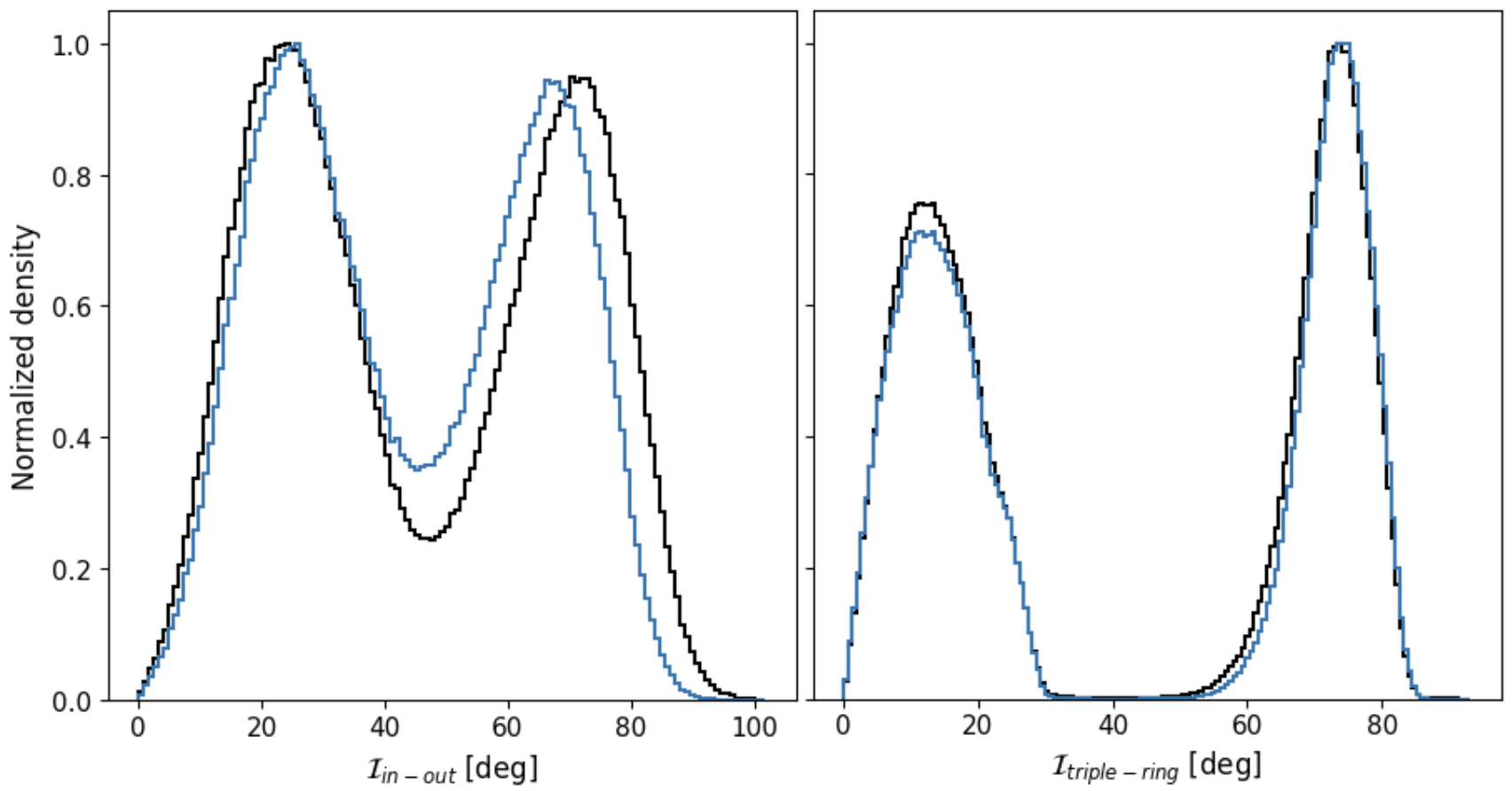}
\caption{Posterior distributions of the misalignment angles for Cases~A and B, shown in black and blue, respectively. Left: Relative inclination between the angular momentum vectors of the inner and outer orbital planes, $\mathcal{I}_{\mathrm{in-out}}$. Right: Relative inclination between the total angular momentum vector of the triple and that of the circumtriple ring, $\mathcal{I}_{\mathrm{triple-ring}}$. These panels highlight the distinct orientation families that remain compatible with the current astrometric constraints.}
\label{fig:mutual_inclinations}
\end{figure*}

Figure~\ref{fig:mutual_inclinations} shows the posterior distributions of $\mathcal{I}_{\mathrm{in-out}}$ and $\mathcal{I}_{\mathrm{triple-ring}}$, while Table~\ref{tab:ggtau-mutual-inclination-peaks} summarizes their main modes for Cases~A and B. The current astrometric constraints permit two distinct orientation families relative to the circumtriple ring: a relatively aligned family centered on $13^\circ$ and a strongly misaligned family centered on $73^\circ$. The two configurations adopted by \citet{Toci2024} for their hydrodynamical simulations both belong to the relatively aligned family. Their compact configuration, with a triple--ring relative inclination of $0^\circ$, has $P_{\rm out}=164$~yr, $a_{\rm out}=34$~au, and $e_{\rm out}=0.28$, and is compatible with the compact solutions allowed by our posterior. Their wider configuration, with a triple--ring relative inclination of $30^\circ$, has $P_{\rm out}=384$~yr, $a_{\rm out}=60$~au, and $e_{\rm out}=0.45$. Although this configuration remains admissible, it is less favored than the more compact and moderately eccentric solutions centered on $a_{\rm out}\simeq43$~au and $e_{\rm out}\simeq0.23$. The strongly misaligned triple--ring family identified here is not represented among the configurations simulated by \citet{Toci2024}. 

{\renewcommand{\arraystretch}{1.4}
\begin{table}[ht!]
\setlength{\tabcolsep}{9.8pt}
\centering
\caption{Summary of the main modes of the inferred misalignment-angle distributions.}
\label{tab:ggtau-mutual-inclination-peaks}
\begin{tabular}{@{}l c r@{\hspace{+0.4em}}l c r@{\hspace{+0.4em}}l r@{\hspace{+0.4em}}l c r@{\hspace{+0.4em}}l@{}}
\hline\hline
Angle & Unit & \multicolumn{5}{c}{Case~A} & \multicolumn{5}{c}{Case~B} \\
\hline
$\mathcal{I}_{\mathrm{in-out}}$
& deg
& $25$ & ${}^{+11}_{-10}$
& $\mkern-25mu-\mkern-25mu$
& $70$ & ${}^{+9}_{-11}$
& $26$ & ${}^{+10}_{-10}$
& $\mkern-25mu-\mkern-25mu$
& $65$ & ${}^{+9}_{-11}$ \\
$\mathcal{I}_{\mathrm{triple-ring}}$
& deg
& $13$ & ${}^{+7}_{-6}$
& $\mkern-25mu|\mkern-25mu$
& $73$ & ${}^{+5}_{-6}$
& $13$ & ${}^{+7}_{-6}$
& $\mkern-25mu|\mkern-25mu$
& $74$ & ${}^{+5}_{-5}$ \\
\hline
\end{tabular}
\tablefoot{The values listed here are given as posterior medians with 16th--84th percentile intervals. The bimodalities of the two angles are only weakly correlated. The misalignment angles were computed directly from the posterior samples and therefore retain the correlations between the orbital parameters.}
\end{table}
}

The relative orientation of the inner and wide stellar orbits also remains degenerate. The posterior distribution of $\mathcal{I}_{\mathrm{in-out}}$ has two main modes near $25^\circ$ and $70^\circ$, corresponding to moderate and strong mutual inclinations, respectively, but also retains admissible configurations at intermediate inclinations. The orientation families associated with $\mathcal{I}_{\mathrm{in-out}}$ and $\mathcal{I}_{\mathrm{triple-ring}}$ are only weakly correlated.

The ring-center constraint slightly narrows both distributions without altering their overall structure. Its impact, however, differs between the two angles. The $\mathcal{I}_{\mathrm{triple-ring}}$ distribution is only weakly affected because this quantity depends on the direction of the total angular momentum of the triple, which is dominated by the wide orbit and is therefore only weakly sensitive to the small changes induced in the inner orbit. By contrast, $\mathcal{I}_{\mathrm{in-out}}$ directly measures the relative orientation of the inner and outer orbits and is thus more sensitive to the modest reweighting of the inner-orbit configurations.

\section{Discussion}
\label{sec:discussion}

The simultaneous fit establishes a self-consistent range of orbital architectures and mass partitions for GG~Tau~$A$, while leaving several geometrical and observational degeneracies unresolved. We first discuss the dynamical implications of the orientation families identified above, before considering the limitations of the present analysis and the observations and modeling needed to refine the solution.

\subsection{Dynamical implications}

The modes of the two mutual-inclination distributions have important dynamical implications. Each angle identifies configurations with distinct secular behavior. As shown in Fig.~\ref{fig:mutual_inclinations}, the solutions that remain compatible with the astrometric constraints alone include configurations with high $\mathcal{I}_{\mathrm{in-out}}$, represented by a mode near $70^\circ$, in which inclination-driven secular effects, potentially up to Kozai-Lidov-type behavior, may become important within the triple itself \citep{Kozai1962,Lidov1962,Naoz2016}. 

They also include configurations with high $\mathcal{I}_{\mathrm{triple-ring}}$, represented by a mode near $73^\circ$, which may favor analogous inclination-eccentricity exchange in the circumtriple disk \citep{Martin2014}. Such highly misaligned, polar configurations, in which the disk angular momentum becomes nearly perpendicular to the orbital plane that dominates its forcing, have already been discussed and constitute dynamically viable solutions \citep{Lubow2018,Martin2019,Lepp2025}. In the specific case of GG~Tau~$A$, a misaligned disk has already been identified as a plausible way to reconcile the observed cavity size with the stellar orbit \citep{Aly2018}. A dedicated dynamical study will therefore be required to determine which of these orientation families can simultaneously reproduce the orbital architecture inferred here and the observed properties of the disk.

\subsection{Limitations and prospects}

These results and their dynamical implications should nevertheless be interpreted in light of several limitations of the present analysis. The first arises from the still incomplete observational constraints on the orbital architecture. Because the analysis relies exclusively on relative astrometry, a residual degeneracy remains in the orientation of the system, particularly with respect to the line of nodes. Breaking this degeneracy will require radial-velocity measurements, and even a single measurement obtained at a favorable orbital phase could already be sufficient to discriminate between the two remaining admissible branches. 

\begin{figure*}[htbp]
  \centering
  \sidecaption
  \includegraphics[width=0.85\textwidth]{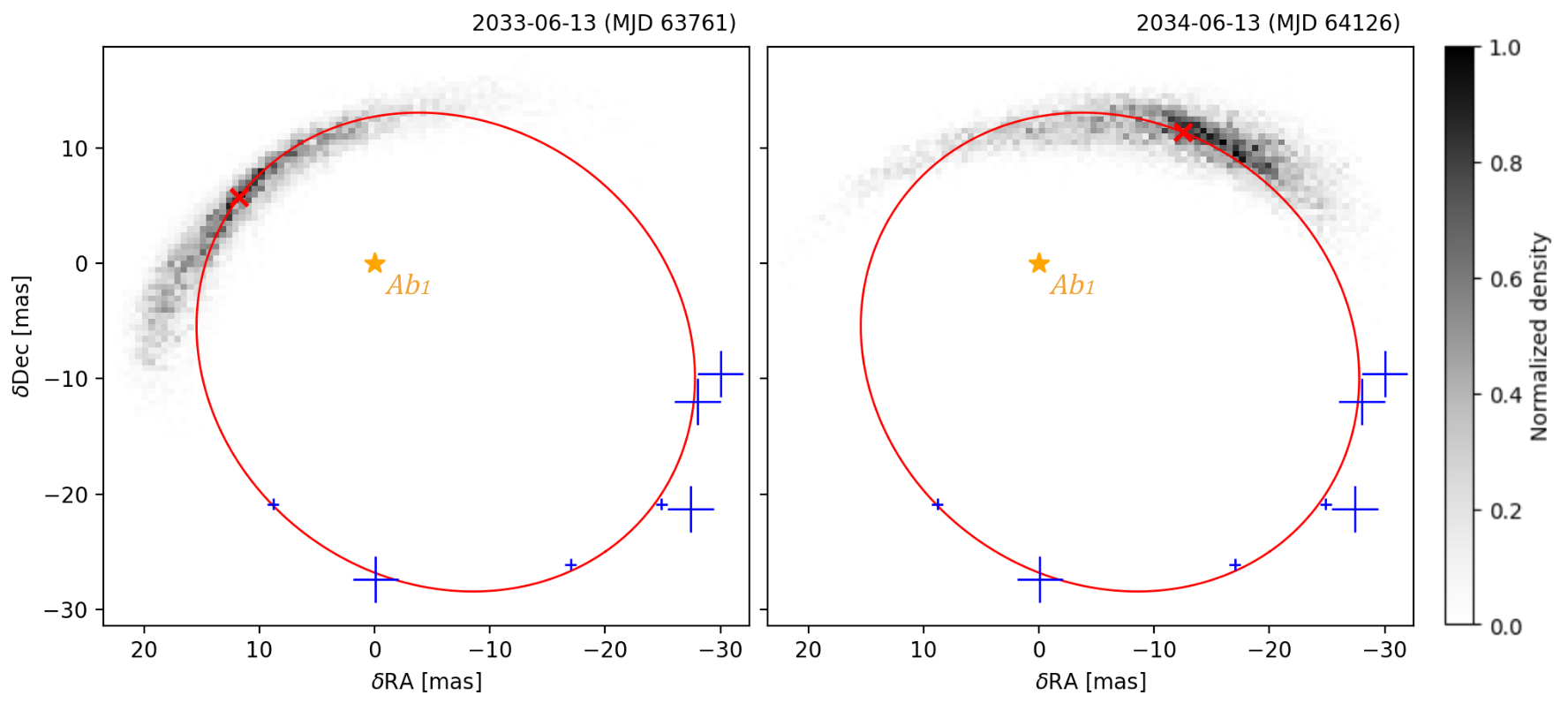}
  \caption{Predicted sky-plane positions of $Ab_2$ relative to $Ab_1$ at two future observing dates for Case~A. Case~B is very similar and is therefore not shown. The grayscale indicates the posterior probability density derived from the orbital fit, with darker regions corresponding to higher probability density. The blue crosses show the astrometric measurements currently available, while the red curve shows the projected orbit of the best-fit solution together with the predicted position of $Ab_2$ at each date.}
  \label{fig:ggtau-ab2-prediction-dates}
\end{figure*}

In addition, the current astrometric coverage of the wide orbit remains limited, so that several of its parameters are still only weakly constrained. A more general tightening of the constraints on the orbital architecture will therefore require a longer astrometric baseline for the wide orbit, together with new astrometric measurements of the inner subsystem obtained at judiciously chosen orbital phases. Figure~\ref{fig:ggtau-ab2-prediction-dates} illustrates this point for the inner orbit by showing, at two future epochs, the predicted position distribution of $Ab_2$ relative to $Ab_1$ in the plane of the sky based on the posterior orbital solutions. These dates are of particular interest because they lie close to periastron passage, a particularly discriminating configuration whose observational window nevertheless remains brief. The next constraining windows will not reopen until 2033 and 2034, after roughly one full orbital period of the inner binary.

A second limitation concerns the reconstruction of the photocenter motion of the $Ab_1$--$Ab_2$ subsystem. This step relies on several methodological assumptions. First, the analysis is restricted to $K$-band measurements, for which the flux ratio between $Ab_2$ and $Ab_1$ is best constrained and appears unvariable. Second, the conversion from $Aa$--$Ab_\mathrm{pc}$ to $Aa$--$Ab_1$ astrometry requires an intermediate solution for the inner binary in order to estimate, at each epoch, the displacement of $Ab_\mathrm{pc}$ relative to $Ab_1$. Although this reconstruction is well justified by the quality of the inner astrometry, it nevertheless introduces an additional layer of modeling and relies on a simplified propagation of the associated uncertainties.

Two further limitations arise from the disk-based constraints. First, the inference of the individual stellar masses remains partly anchored by the external prior on the total mass derived from disk kinematics, while both the physical scale of the system and the stellar masses depend on the adopted distance. We adopted 145~pc, consistently with recent studies of GG~Tau~$A$ and with the mean distance of the L1551 group, to which the system belongs \citep{Galli2019,Toci2024,Duchene2024}. This choice is consistent with Gaia data release 2 (DR2) and data release 3 (DR3) measurements of the GG~Tau~$B$ subsystem, which suggest distances ranging from approximately 145 to 150~pc \citep{Gaia2016,Gaia2018,Gaia2023}. These estimates should nevertheless be treated with caution because GG~Tau~$B$ is itself a binary, although its parallax is expected to be less affected by multiplicity than that of GG~Tau~$A$.

With relative astrometry alone, physical orbital dimensions scale linearly with distance and dynamical masses with the cube of the distance, whereas angular geometry, periods, and mass ratios remain unchanged. The total-mass prior derived from disk kinematics, however, scales only linearly with distance through the Keplerian rotation curve. A revision of the adopted distance would therefore require a new fit with a consistently rescaled mass prior. For instance, adopting 150~pc rather than 145~pc would increase the mass scale inferred from relative astrometry alone by 11\%, while increasing the disk-based total-mass prior by only 3\%. The individual masses could consequently not all increase uniformly by 11\%. Because the disk-based prior anchors the total mass, the masses of $Aa$ and $Ab_1$ would likely increase only modestly, while a larger relative increase could be accommodated for $Ab_2$.

Second, the ring-center constraint relies on the assumption that the circumtriple ring is globally axisymmetric, such that its geometrical center traces the projected barycenter of the triple. This assumption may be oversimplified, since hydrodynamical simulations show that circumbinary disks can develop eccentric and non-axisymmetric structures \citep{Ragusa2020}. Moreover, similar effects may arise in circumtriple disks. The ring-center measurement is referenced to the centroid of the compact continuum emission associated with $Aa$, which is assumed to coincide with the stellar position. Any systematic offset between this dust-emission centroid and the stellar photosphere is not included in the uncertainty budget.

For these reasons, we applied the ring-center information through posterior reweighting rather than including it directly in a fully joint fit. This approach is appropriate for an observable that is statistically independent of the stellar astrometry and remains subject to systematic uncertainties. Additional independent measurements of the ring-center position would allow the robustness of this geometrical constraint to be assessed and could justify its direct incorporation into the fit as an additional prior on the stellar mass ratios, analogously to the prior adopted for the total system mass. For the current single measurement, however, the post-processing approach remains preferable because it makes the impact of this relatively uncertain constraint explicit.

\section{Conclusions}
\label{sec:conclusion}

We presented the first simultaneous hierarchical fit of the GG~Tau~$A$ triple system, in which the inner $Ab_2$--$Ab_1$ orbit and the wide $Aa$--$Ab_\mathrm{cm}$ orbit are solved jointly rather than through separate fits. A key step of the analysis is the conversion of the historical $Aa$--$Ab_\mathrm{pc}$ astrometry into $Aa$--$Ab_1$ measurements, which places all astrometric observations in a common reference frame and allows the coupling between the two orbital levels to be propagated consistently. This joint treatment yields a self-consistent orbital architecture for the system and direct constraints on the individual stellar masses.

The additional wide-orbit astrometric epoch only marginally sharpens the posterior distributions, which remains consistent with the still limited coverage of the outer orbit. By contrast, the two disk-based constraints have distinct effects on the solution. The mass enclosed within the disk, as inferred from its kinematics, strongly constrains the total mass of the triple, whereas the position of the ring center constrains that of the triple barycenter and, consequently, the distribution of mass among the three components. Taken together, these constraints do not qualitatively modify the preferred orbital architecture, but they significantly restrict the range of admissible barycentric configurations and thereby tighten the posterior distributions of the individual stellar masses. In particular, they favor solutions in which $Aa$ is slightly more massive than $Ab_1$, in better agreement with the spectroscopic analysis of \citet{White1999,Hartigan2003}. The final solution is therefore consistent both with the inner and wide astrometry and with the kinematic and geometrical constraints derived from the circumtriple disk. 

Another notable result is that the inferred mass ratio of the inner $Ab$ binary is very close to the observed $K$-band flux ratio. This provides retrospective justification for the use of the photocenter as a proxy for its barycenter in the wide-orbit astrometric analysis of \citet{Toci2024}.

More generally, the methodology developed here is applicable to hierarchical multiple systems over a broad range of orbital configurations. Its constraining power is expected to be particularly strong in compact systems, where a larger fraction of the orbital motion can be covered over a given observational baseline, potentially yielding tighter constraints on the system architecture and on the individual dynamical masses.

For GG~Tau~$A$, a third disk-based constraint is provided by the observed inner edge of the circumtriple disk, which will be examined in a forthcoming companion paper. Unlike the total-mass and ring-center constraints considered here, the cavity size probes the cumulative dynamical response of the disk to the hierarchical triple rather than a purely kinematic or geometrical property of the system. Interpreting this observable will therefore require dedicated dynamical modeling of the distinct orbital families identified in this work. By confronting these models with the observed cavity size and disk morphology, the companion study will assess whether the remaining degeneracies can be resolved and whether a coherent dynamical picture of the system can be established.

\begin{acknowledgements} 
We are grateful to an anonymous referee for feedback that helped us improve this manuscript. All computations presented in this paper were performed using the GRICAD infrastructure (\url{https://gricad.univ-grenoble-alpes.fr}), which is supported by Grenoble research communities. 
\end{acknowledgements}

\bibliographystyle{aa}
\bibliography{bibli}

@ARTICLE{Toci2024,
       author = {{Toci}, Claudia and {Ceppi}, Simone and {Cuello}, Nicol{\'a}s and {Duch{\^e}ne}, Gaspard and {Ragusa}, Enrico and {Lodato}, Giuseppe and {Farina}, Francesca and {M{\'e}nard}, Fran{\c{c}}ois and {Aly}, Hossam},
        title = "{Orbital dynamics in the GG Tau A system: Investigating its enigmatic disc}",
      journal = {\aap},
         year = 2024,
        month = aug,
       volume = {688},
          eid = {A102},
        pages = {A102},
          doi = {10.1051/0004-6361/202348470},
archivePrefix = {arXiv},
       eprint = {2404.07565},
 primaryClass = {astro-ph.EP},
       adsurl = {https://ui.adsabs.harvard.edu/abs/2024A&A...688A.102T}
}

@ARTICLE{Duchene2024,
       author = {{Duch{\^e}ne}, Gaspard and {LeBouquin}, Jean-Baptiste and {M{\'e}nard}, Fran{\c{c}}ois and {Cuello}, Nicol{\'a}s and {Toci}, Claudia and {Langlois}, Maud},
        title = "{Full orbital solutions in pre-main sequence high-order multiple systems: GG Tau Ab and UX Tau B}",
      journal = {\aap},
         year = 2024,
        month = jun,
       volume = {686},
          eid = {A188},
        pages = {A188},
          doi = {10.1051/0004-6361/202348755},
archivePrefix = {arXiv},
       eprint = {2404.02469},
 primaryClass = {astro-ph.SR},
       adsurl = {https://ui.adsabs.harvard.edu/abs/2024A&A...686A.188D}
}

@ARTICLE{Tokovinin2021,
       author = {{Tokovinin}, Andrei},
        title = "{Architecture of Hierarchical Stellar Systems and Their Formation}",
      journal = {Universe},
         year = 2021,
        month = sep,
       volume = {7},
       number = {9},
          eid = {352},
        pages = {352},
          doi = {10.3390/universe7090352},
archivePrefix = {arXiv},
       eprint = {2109.09118},
 primaryClass = {astro-ph.SR},
       adsurl = {https://ui.adsabs.harvard.edu/abs/2021Univ....7..352T}
}

@ARTICLE{Leinert1993,
       author = {{Leinert}, Ch. and {Zinnecker}, H. and {Weitzel}, N. and {Christou}, J. and {Ridgway}, S.~T. and {Jameson}, R. and {Haas}, M. and {Lenzen}, R.},
        title = "{A systematic search for young binaries in Taurus.}",
      journal = {\aap},
         year = 1993,
        month = oct,
       volume = {278},
        pages = {129-149},
       adsurl = {https://ui.adsabs.harvard.edu/abs/1993A&A...278..129L}
}

@ARTICLE{Silber2000,
       author = {{Silber}, Joel and {Gledhill}, Tim and {Duch{\^e}ne}, Gaspard and {M{\'e}nard}, Fran{\c{c}}ois},
        title = "{Near-Infrared Imaging Polarimetry of the GG Tauri Circumbinary Ring}",
      journal = {\apjl},
         year = 2000,
        month = jun,
       volume = {536},
       number = {2},
        pages = {L89-L92},
          doi = {10.1086/312731},
archivePrefix = {arXiv},
       eprint = {astro-ph/0005303},
 primaryClass = {astro-ph},
       adsurl = {https://ui.adsabs.harvard.edu/abs/2000ApJ...536L..89S}
}

@ARTICLE{Beck2012,
       author = {{Beck}, Tracy L. and {Bary}, Jeffrey S. and {Dutrey}, Anne and {Pi{\'e}tu}, Vincent and {Guilloteau}, St{\'e}phane and {Lubow}, S.~H. and {Simon}, M.},
        title = "{Circumbinary Gas Accretion onto a Central Binary: Infrared Molecular Hydrogen Emission from GG Tau A}",
      journal = {\apj},
         year = 2012,
        month = jul,
       volume = {754},
       number = {1},
          eid = {72},
        pages = {72},
          doi = {10.1088/0004-637X/754/1/72},
archivePrefix = {arXiv},
       eprint = {1205.1526},
 primaryClass = {astro-ph.SR},
       adsurl = {https://ui.adsabs.harvard.edu/abs/2012ApJ...754...72B}
}

@ARTICLE{Krist2002,
       author = {{Krist}, John E. and {Stapelfeldt}, Karl R. and {Watson}, Alan M.},
        title = "{Hubble Space Telescope/WFPC2 Images of the GG Tauri Circumbinary Disk}",
      journal = {\apj},
         year = 2002,
        month = may,
       volume = {570},
       number = {2},
        pages = {785-792},
          doi = {10.1086/339777},
archivePrefix = {arXiv},
       eprint = {astro-ph/0201415},
 primaryClass = {astro-ph},
       adsurl = {https://ui.adsabs.harvard.edu/abs/2002ApJ...570..785K}
}

@ARTICLE{Duchene2004,
       author = {{Duch{\^e}ne}, G. and {McCabe}, C. and {Ghez}, A.~M. and {Macintosh}, B.~A.},
        title = "{A Multiwavelength Scattered Light Analysis of the Dust Grain Population in the GG Tauri Circumbinary Ring}",
      journal = {\apj},
         year = 2004,
        month = may,
       volume = {606},
       number = {2},
        pages = {969-982},
          doi = {10.1086/383126},
archivePrefix = {arXiv},
       eprint = {astro-ph/0401560},
 primaryClass = {astro-ph},
       adsurl = {https://ui.adsabs.harvard.edu/abs/2004ApJ...606..969D}
}

@ARTICLE{Andrews2014,
       author = {{Andrews}, Sean M. and {Chandler}, Claire J. and {Isella}, Andrea and {Birnstiel}, T. and {Rosenfeld}, K.~A. and {Wilner}, D.~J. and {P{\'e}rez}, L.~M. and {Ricci}, L. and {Carpenter}, J.~M. and {Calvet}, N. and {Corder}, S.~A. and {Deller}, A.~T. and {Dullemond}, C.~P. and {Greaves}, J.~S. and {Harris}, R.~J. and {Henning}, Th. and {Kwon}, W. and {Lazio}, J. and {Linz}, H. and {Mundy}, L.~G. and {Sargent}, A.~I. and {Storm}, S. and {Testi}, L.},
        title = "{Resolved Multifrequency Radio Observations of GG Tau}",
      journal = {\apj},
         year = 2014,
        month = jun,
       volume = {787},
       number = {2},
          eid = {148},
        pages = {148},
          doi = {10.1088/0004-637X/787/2/148},
archivePrefix = {arXiv},
       eprint = {1404.5652},
 primaryClass = {astro-ph.SR},
       adsurl = {https://ui.adsabs.harvard.edu/abs/2014ApJ...787..148A}
}

@ARTICLE{Cazzoletti2017,
       author = {{Cazzoletti}, P. and {Ricci}, L. and {Birnstiel}, T. and {Lodato}, G.},
        title = "{Testing dust trapping in the circumbinary disk around GG Tauri A}",
      journal = {\aap},
         year = 2017,
        month = mar,
       volume = {599},
          eid = {A102},
        pages = {A102},
          doi = {10.1051/0004-6361/201629721},
archivePrefix = {arXiv},
       eprint = {1610.08381},
 primaryClass = {astro-ph.SR},
       adsurl = {https://ui.adsabs.harvard.edu/abs/2017A&A...599A.102C}
}

@ARTICLE{Keppler2020,
       author = {{Keppler}, M. and {Penzlin}, A. and {Benisty}, M. and {van Boekel}, R. and {Henning}, T. and {van Holstein}, R.~G. and {Kley}, W. and {Garufi}, A. and {Ginski}, C. and {Brandner}, W. and {Bertrang}, G.~H.-M. and {Boccaletti}, A. and {de Boer}, J. and {Bonavita}, M. and {Brown Sevilla}, S. and {Chauvin}, G. and {Dominik}, C. and {Janson}, M. and {Langlois}, M. and {Lodato}, G. and {Maire}, A.-L. and {M{\'e}nard}, F. and {Pantin}, E. and {Pinte}, C. and {Stolker}, T. and {Szul{\'a}gyi}, J. and {Thebault}, P. and {Villenave}, M. and {Zurlo}, A. and {Rabou}, P. and {Feautrier}, P. and {Feldt}, M. and {Madec}, F. and {Wildi}, F.},
        title = "{Gap, shadows, spirals, and streamers: SPHERE observations of binary-disk interactions in GG Tauri A}",
      journal = {\aap},
         year = 2020,
        month = jul,
       volume = {639},
          eid = {A62},
        pages = {A62},
          doi = {10.1051/0004-6361/202038032},
archivePrefix = {arXiv},
       eprint = {2005.09037},
 primaryClass = {astro-ph.SR},
       adsurl = {https://ui.adsabs.harvard.edu/abs/2020A&A...639A..62K}
}

@ARTICLE{DucheneKraus2013,
       author = {{Duch{\^e}ne}, Gaspard and {Kraus}, Adam},
        title = "{Stellar Multiplicity}",
      journal = {\araa},
         year = 2013,
        month = aug,
       volume = {51},
       number = {1},
        pages = {269-310},
          doi = {10.1146/annurev-astro-081710-102602},
archivePrefix = {arXiv},
       eprint = {1303.3028},
 primaryClass = {astro-ph.SR},
       adsurl = {https://ui.adsabs.harvard.edu/abs/2013ARA&A..51..269D}
}

@INPROCEEDINGS{Offner2023,
       author = {{Offner}, S.~S.~R. and {Moe}, M. and {Kratter}, K.~M. and {Sadavoy}, S.~I. and {Jensen}, E.~L.~N. and {Tobin}, J.~J.},
        title = "{The Origin and Evolution of Multiple Star Systems}",
    booktitle = {Protostars and Planets VII},
         year = 2023,
       editor = {{Inutsuka}, S. and {Aikawa}, Y. and {Muto}, T. and {Tomida}, K. and {Tamura}, M.},
       series = {Astronomical Society of the Pacific Conference Series},
       volume = {534},
        month = jul,
        pages = {275},
          doi = {10.48550/arXiv.2203.10066},
archivePrefix = {arXiv},
       eprint = {2203.10066},
 primaryClass = {astro-ph.SR},
       adsurl = {https://ui.adsabs.harvard.edu/abs/2023ASPC..534..275O}
}

@ARTICLE{ArtymowiczLubow1994,
       author = {{Artymowicz}, Pawel and {Lubow}, Stephen H.},
        title = "{Dynamics of Binary-Disk Interaction. I. Resonances and Disk Gap Sizes}",
      journal = {\apj},
         year = 1994,
        month = feb,
       volume = {421},
        pages = {651},
          doi = {10.1086/173679},
       adsurl = {https://ui.adsabs.harvard.edu/abs/1994ApJ...421..651A}
}

@ARTICLE{Ceppi2022,
       author = {{Ceppi}, Simone and {Cuello}, Nicol{\'a}s and {Lodato}, Giuseppe and {Clarke}, Cathie and {Toci}, Claudia and {Price}, Daniel J.},
        title = "{Accretion rates in hierarchical triple systems with discs}",
      journal = {\mnras},
         year = 2022,
        month = jul,
       volume = {514},
       number = {1},
        pages = {906-919},
          doi = {10.1093/mnras/stac1390},
archivePrefix = {arXiv},
       eprint = {2205.08784},
 primaryClass = {astro-ph.SR},
       adsurl = {https://ui.adsabs.harvard.edu/abs/2022MNRAS.514..906C}
}

@ARTICLE{Ceppi2023,
       author = {{Ceppi}, Simone and {Longarini}, Cristiano and {Lodato}, Giuseppe and {Cuello}, Nicol{\'a}s and {Lubow}, Stephen H.},
        title = "{Precession and polar alignment of accretion discs in triple (or multiple) stellar systems}",
      journal = {\mnras},
         year = 2023,
        month = apr,
       volume = {520},
       number = {4},
        pages = {5817-5827},
          doi = {10.1093/mnras/stad444},
archivePrefix = {arXiv},
       eprint = {2302.03411},
 primaryClass = {astro-ph.EP},
       adsurl = {https://ui.adsabs.harvard.edu/abs/2023MNRAS.520.5817C}
}

@ARTICLE{BeustDutrey2005,
       author = {{Beust}, H. and {Dutrey}, A.},
        title = "{Dynamics of the young multiple system <ASTROBJ>GG Tauri</ASTROBJ>. I. Orbital fits and inner edge of the circumbinary disk of <ASTROBJ>GG Tau</ASTROBJ> A}",
      journal = {\aap},
         year = 2005,
        month = aug,
       volume = {439},
       number = {2},
        pages = {585-594},
          doi = {10.1051/0004-6361:20042441},
       adsurl = {https://ui.adsabs.harvard.edu/abs/2005A&A...439..585B}
}

@ARTICLE{Koehler2011,
       author = {{K{\"o}hler}, R.},
        title = "{The orbit of GG Tauri A}",
      journal = {\aap},
         year = 2011,
        month = jun,
       volume = {530},
          eid = {A126},
        pages = {A126},
          doi = {10.1051/0004-6361/201016327},
archivePrefix = {arXiv},
       eprint = {1104.2245},
 primaryClass = {astro-ph.SR},
       adsurl = {https://ui.adsabs.harvard.edu/abs/2011A&A...530A.126K}
}

@ARTICLE{NelsonMarzari2016,
       author = {{Nelson}, Andrew F. and {Marzari}, F.},
        title = "{Dynamics of Circumstellar Disks. III. The Case of GG Tau A}",
      journal = {\apj},
         year = 2016,
        month = aug,
       volume = {827},
       number = {2},
          eid = {93},
        pages = {93},
          doi = {10.3847/0004-637X/827/2/93},
archivePrefix = {arXiv},
       eprint = {1605.02764},
 primaryClass = {astro-ph.SR},
       adsurl = {https://ui.adsabs.harvard.edu/abs/2016ApJ...827...93N}
}

@ARTICLE{Galli2019,
       author = {{Galli}, P.~A.~B. and {Loinard}, L. and {Bouy}, H. and {Sarro}, L.~M. and {Ortiz-Le{\'o}n}, G.~N. and {Dzib}, S.~A. and {Olivares}, J. and {Heyer}, M. and {Hernandez}, J. and {Rom{\'a}n-Z{\'u}{\~n}iga}, C. and {Kounkel}, M. and {Covey}, K.},
        title = "{Structure and kinematics of the Taurus star-forming region from Gaia-DR2 and VLBI astrometry}",
      journal = {\aap},
         year = 2019,
        month = oct,
       volume = {630},
          eid = {A137},
        pages = {A137},
          doi = {10.1051/0004-6361/201935928},
archivePrefix = {arXiv},
       eprint = {1909.01118},
 primaryClass = {astro-ph.SR},
       adsurl = {https://ui.adsabs.harvard.edu/abs/2019A&A...630A.137G}
}

@ARTICLE{Halbwachs2023,
       author = {{Halbwachs}, Jean-Louis and {Pourbaix}, Dimitri and {Arenou}, Fr{\'e}d{\'e}ric and {Galluccio}, Laurent and {Guillout}, Patrick and {Bauchet}, Nathalie and {Marchal}, Olivier and {Sadowski}, Gilles and {Teyssier}, David},
        title = "{Gaia Data Release 3. Astrometric binary star processing}",
      journal = {\aap},
         year = 2023,
        month = jun,
       volume = {674},
          eid = {A9},
        pages = {A9},
          doi = {10.1051/0004-6361/202243969},
archivePrefix = {arXiv},
       eprint = {2206.05726},
 primaryClass = {astro-ph.SR},
       adsurl = {https://ui.adsabs.harvard.edu/abs/2023A&A...674A...9H}
}

@ARTICLE{Belokurov2020,
       author = {{Belokurov}, Vasily and {Penoyre}, Zephyr and {Oh}, Semyeong and {Iorio}, Giuliano and {Hodgkin}, Simon and {Evans}, N. Wyn and {Everall}, Andrew and {Koposov}, Sergey E. and {Tout}, Christopher A. and {Izzard}, Robert and {Clarke}, Cathie J. and {Brown}, Anthony G.~A.},
        title = "{Unresolved stellar companions with Gaia DR2 astrometry}",
      journal = {\mnras},
         year = 2020,
        month = aug,
       volume = {496},
       number = {2},
        pages = {1922-1940},
          doi = {10.1093/mnras/staa1522},
archivePrefix = {arXiv},
       eprint = {2003.05467},
 primaryClass = {astro-ph.SR},
       adsurl = {https://ui.adsabs.harvard.edu/abs/2020MNRAS.496.1922B}
}

@ARTICLE{Ford2006,
       author = {{Ford}, Eric B.},
        title = "{Improving the Efficiency of Markov Chain Monte Carlo for Analyzing the Orbits of Extrasolar Planets}",
      journal = {\apj},
         year = 2006,
        month = may,
       volume = {642},
       number = {1},
        pages = {505-522},
          doi = {10.1086/500802},
archivePrefix = {arXiv},
       eprint = {astro-ph/0512634},
 primaryClass = {astro-ph},
       adsurl = {https://ui.adsabs.harvard.edu/abs/2006ApJ...642..505F}
}

@ARTICLE{Ford2005,
       author = {{Ford}, Eric B.},
        title = "{Quantifying the Uncertainty in the Orbits of Extrasolar Planets}",
      journal = {\aj},
         year = 2005,
        month = mar,
       volume = {129},
       number = {3},
        pages = {1706-1717},
          doi = {10.1086/427962},
archivePrefix = {arXiv},
       eprint = {astro-ph/0305441},
 primaryClass = {astro-ph},
       adsurl = {https://ui.adsabs.harvard.edu/abs/2005AJ....129.1706F}
}

@ARTICLE{Phuong2018,
       author = {{Phuong}, Nguyen Thi and {Diep}, Pham Ngoc and {Dutrey}, Anne and {Chapillon}, Edwige and {Darriulat}, Pierre and {Guilloteau}, St{\'e}phane and {Hoai}, Do Thi and {Tuyet Nhung}, Pham and {Tang}, Ya-Wen and {Thao}, Nguyen Thi and {Tuan-Anh}, Pham},
        title = "{Morphology of the $^{13}$CO(3-2) millimetre emission across the gas disc surrounding the triple protostar GG Tau A using ALMA observations}",
      journal = {Research in Astronomy and Astrophysics},
         year = 2018,
        month = mar,
       volume = {18},
       number = {3},
          eid = {031},
        pages = {031},
          doi = {10.1088/1674-4527/18/3/31},
archivePrefix = {arXiv},
       eprint = {1801.00861},
 primaryClass = {astro-ph.SR},
       adsurl = {https://ui.adsabs.harvard.edu/abs/2018RAA....18...31P}
}

@ARTICLE{Phuong2020,
       author = {{Phuong}, N.~T. and {Dutrey}, A. and {Diep}, P.~N. and {Guilloteau}, S. and {Chapillon}, E. and {Di Folco}, E. and {Tang}, Y.-W. and {Pi{\'e}tu}, V. and {Bary}, J. and {Beck}, T. and {Hersant}, F. and {Hoai}, D.~T. and {Hur{\'e}}, J.~M. and {Nhung}, P.~T. and {Pierens}, A. and {Tuan-Anh}, P.},
        title = "{GG Tauri A: gas properties and dynamics from the cavity to the outer disk}",
      journal = {\aap},
         year = 2020,
        month = mar,
       volume = {635},
          eid = {A12},
        pages = {A12},
          doi = {10.1051/0004-6361/201936173},
archivePrefix = {arXiv},
       eprint = {2001.08147},
 primaryClass = {astro-ph.SR},
       adsurl = {https://ui.adsabs.harvard.edu/abs/2020A&A...635A..12P}
}

@ARTICLE{Dutrey1994,
       author = {{Dutrey}, A. and {Guilloteau}, S. and {Simon}, M.},
        title = "{Images of the GG Tauri rotating ring}",
      journal = {\aap},
         year = 1994,
        month = jun,
       volume = {286},
        pages = {149-159},
       adsurl = {https://ui.adsabs.harvard.edu/abs/1994A&A...286..149D}
}

@ARTICLE{Guilloteau1999,
       author = {{Guilloteau}, S. and {Dutrey}, A. and {Simon}, M.},
        title = "{GG Tauri: the ring world}",
      journal = {\aap},
         year = 1999,
        month = aug,
       volume = {348},
        pages = {570-578},
       adsurl = {https://ui.adsabs.harvard.edu/abs/1999A&A...348..570G}
}

@ARTICLE{Dutrey2014,
       author = {{Dutrey}, Anne and {di Folco}, Emmanuel and {Guilloteau}, St{\'e}phane and {Boehler}, Yann and {Bary}, Jeff and {Beck}, Tracy and {Beust}, Herv{\'e} and {Chapillon}, Edwige and {Gueth}, Fred{\'e}ric and {Hur{\'e}}, Jean-Marc and {Pierens}, Arnaud and {Pi{\'e}tu}, Vincent and {Simon}, Michal and {Tang}, Ya-Wen},
        title = "{Possible planet formation in the young, low-mass, multiple stellar system GG Tau A}",
      journal = {\nat},
         year = 2014,
        month = oct,
       volume = {514},
       number = {7524},
        pages = {600-602},
          doi = {10.1038/nature13822},
       adsurl = {https://ui.adsabs.harvard.edu/abs/2014Natur.514..600D}
}

@ARTICLE{DiFolco2014,
       author = {{Di Folco}, E. and {Dutrey}, A. and {Le Bouquin}, J.-B. and {Lacour}, S. and {Berger}, J.-P. and {K{\"o}hler}, R. and {Guilloteau}, S. and {Pi{\'e}tu}, V. and {Bary}, J. and {Beck}, T. and {Beust}, H. and {Pantin}, E.},
        title = "{GG Tauri: the fifth element}",
      journal = {\aap},
         year = 2014,
        month = may,
       volume = {565},
          eid = {L2},
        pages = {L2},
          doi = {10.1051/0004-6361/201423675},
archivePrefix = {arXiv},
       eprint = {1404.2205},
 primaryClass = {astro-ph.SR},
       adsurl = {https://ui.adsabs.harvard.edu/abs/2014A&A...565L...2D}
}

@ARTICLE{Aly2018,
       author = {{Aly}, Hossam and {Lodato}, Giuseppe and {Cazzoletti}, Paolo},
        title = "{On the secular evolution of GG Tau A circumbinary disc: a misaligned disc scenario}",
      journal = {\mnras},
         year = 2018,
        month = nov,
       volume = {480},
       number = {4},
        pages = {4738-4745},
          doi = {10.1093/mnras/sty2179},
archivePrefix = {arXiv},
       eprint = {1809.06383},
 primaryClass = {astro-ph.SR},
       adsurl = {https://ui.adsabs.harvard.edu/abs/2018MNRAS.480.4738A}
}

@ARTICLE{Martin2014,
       author = {{Martin}, Rebecca G. and {Nixon}, Chris and {Lubow}, Stephen H. and {Armitage}, Philip J. and {Price}, Daniel J. and {Do{\u{g}}an}, Suzan and {King}, Andrew},
        title = "{The Kozai-Lidov Mechanism in Hydrodynamical Disks}",
      journal = {\apjl},
         year = 2014,
        month = sep,
       volume = {792},
       number = {2},
          eid = {L33},
        pages = {L33},
          doi = {10.1088/2041-8205/792/2/L33},
archivePrefix = {arXiv},
       eprint = {1409.1226},
 primaryClass = {astro-ph.EP},
       adsurl = {https://ui.adsabs.harvard.edu/abs/2014ApJ...792L..33M}
}

@ARTICLE{Lubow2018,
       author = {{Lubow}, Stephen H. and {Martin}, Rebecca G.},
        title = "{Linear analysis of the evolution of nearly polar low-mass circumbinary discs}",
      journal = {\mnras},
         year = 2018,
        month = jan,
       volume = {473},
       number = {3},
        pages = {3733-3746},
          doi = {10.1093/mnras/stx2643},
archivePrefix = {arXiv},
       eprint = {1710.02233},
 primaryClass = {astro-ph.SR},
       adsurl = {https://ui.adsabs.harvard.edu/abs/2018MNRAS.473.3733L}
}

@ARTICLE{Martin2019,
       author = {{Martin}, Rebecca G. and {Lubow}, Stephen H.},
        title = "{Polar alignment of a protoplanetary disc around an eccentric binary - III. Effect of disc mass}",
      journal = {\mnras},
         year = 2019,
        month = nov,
       volume = {490},
       number = {1},
        pages = {1332-1349},
          doi = {10.1093/mnras/stz2670},
archivePrefix = {arXiv},
       eprint = {1904.11631},
 primaryClass = {astro-ph.EP},
       adsurl = {https://ui.adsabs.harvard.edu/abs/2019MNRAS.490.1332M}
}

@ARTICLE{Kozai1962,
       author = {{Kozai}, Yoshihide},
        title = "{Secular perturbations of asteroids with high inclination and eccentricity}",
      journal = {\aj},
         year = 1962,
        month = nov,
       volume = {67},
        pages = {591-598},
          doi = {10.1086/108790},
       adsurl = {https://ui.adsabs.harvard.edu/abs/1962AJ.....67..591K}
}

@ARTICLE{Lidov1962,
       author = {{Lidov}, M.~L.},
        title = "{The evolution of orbits of artificial satellites of planets under the action of gravitational perturbations of external bodies}",
      journal = {\planss},
         year = 1962,
        month = oct,
       volume = {9},
       number = {10},
        pages = {719-759},
          doi = {10.1016/0032-0633(62)90129-0},
       adsurl = {https://ui.adsabs.harvard.edu/abs/1962P&SS....9..719L}
}

@ARTICLE{Naoz2016,
       author = {{Naoz}, Smadar},
        title = "{The Eccentric Kozai-Lidov Effect and Its Applications}",
      journal = {\araa},
         year = 2016,
        month = sep,
       volume = {54},
        pages = {441-489},
          doi = {10.1146/annurev-astro-081915-023315},
archivePrefix = {arXiv},
       eprint = {1601.07175},
 primaryClass = {astro-ph.EP},
       adsurl = {https://ui.adsabs.harvard.edu/abs/2016ARA&A..54..441N}
}

@ARTICLE{WilliamsCieza2011,
       author = {{Williams}, Jonathan P. and {Cieza}, Lucas A.},
        title = "{Protoplanetary Disks and Their Evolution}",
      journal = {\araa},
         year = 2011,
        month = sep,
       volume = {49},
       number = {1},
        pages = {67-117},
          doi = {10.1146/annurev-astro-081710-102548},
archivePrefix = {arXiv},
       eprint = {1103.0556},
 primaryClass = {astro-ph.GA},
       adsurl = {https://ui.adsabs.harvard.edu/abs/2011ARA&A..49...67W}
}

@ARTICLE{DanbyBurkardt1983,
       author = {{Danby}, J.~M.~A. and {Burkardt}, T.~M.},
        title = "{The Solution of Kepler's Equation - Part One}",
      journal = {Celestial Mechanics},
         year = 1983,
        month = oct,
       volume = {31},
       number = {2},
        pages = {95-107},
          doi = {10.1007/BF01686811},
       adsurl = {https://ui.adsabs.harvard.edu/abs/1983CeMec..31...95D}
}

@ARTICLE{BurkardtDanby1983,
       author = {{Burkardt}, T.~M. and {Danby}, J.~M.~A.},
        title = "{The Solution of Kepler's Equation - Part Two}",
      journal = {Celestial Mechanics},
         year = 1983,
        month = nov,
       volume = {31},
       number = {3},
        pages = {317-328},
          doi = {10.1007/BF01844230},
       adsurl = {https://ui.adsabs.harvard.edu/abs/1983CeMec..31..317B}
}

@ARTICLE{Danby1987,
       author = {{Danby}, J.~M.~A.},
        title = "{The Solution of Kepler's Equations - Part Three}",
      journal = {Celestial Mechanics},
         year = 1987,
        month = sep,
       volume = {40},
       number = {3-4},
        pages = {303-312},
          doi = {10.1007/BF01235847},
       adsurl = {https://ui.adsabs.harvard.edu/abs/1987CeMec..40..303D}
}

@ARTICLE{Beust2016,
       author = {{Beust}, H. and {Bonnefoy}, M. and {Maire}, A.-L. and {Ehrenreich}, D. and {Lagrange}, A.-M. and {Chauvin}, G.},
        title = "{Orbital fitting of imaged planetary companions with high eccentricities and unbound orbits. Their application to Fomalhaut b and PZ Telecopii B}",
      journal = {\aap},
         year = 2016,
        month = mar,
       volume = {587},
          eid = {A89},
        pages = {A89},
          doi = {10.1051/0004-6361/201527388},
archivePrefix = {arXiv},
       eprint = {1512.03596},
 primaryClass = {astro-ph.EP},
       adsurl = {https://ui.adsabs.harvard.edu/abs/2016A&A...587A..89B}
}

@BOOK{Plummer1918,
       author = {{Plummer}, Henry Crozier Keating},
        title = "{An introductory treatise on dynamical astronomy}",
         year = 1918,
       adsurl = {https://ui.adsabs.harvard.edu/abs/1918itda.book.....P}
}

@ARTICLE{Beust2003,
       author = {{Beust}, H.},
        title = "{Symplectic integration of hierarchical stellar systems}",
      journal = {\aap},
         year = 2003,
        month = mar,
       volume = {400},
        pages = {1129-1144},
          doi = {10.1051/0004-6361:20030065},
       adsurl = {https://ui.adsabs.harvard.edu/abs/2003A&A...400.1129B}
}

@ARTICLE{White1999,
       author = {{White}, Russel J. and {Ghez}, A.~M. and {Reid}, I. Neill and {Schultz}, Greg},
        title = "{A Test of Pre-Main-Sequence Evolutionary Models across the Stellar/Substellar Boundary Based on Spectra of the Young Quadruple GG Tauri}",
      journal = {\apj},
         year = 1999,
        month = aug,
       volume = {520},
       number = {2},
        pages = {811-821},
          doi = {10.1086/307494},
archivePrefix = {arXiv},
       eprint = {astro-ph/9902318},
 primaryClass = {astro-ph},
       adsurl = {https://ui.adsabs.harvard.edu/abs/1999ApJ...520..811W}
}

@ARTICLE{Beuzit2019,
       author = {{Beuzit}, J.-L. and {Vigan}, A. and {Mouillet}, D. and {Dohlen}, K. and {Gratton}, R. and {Boccaletti}, A. and {Sauvage}, J.-F. and {Schmid}, H.~M. and {Langlois}, M. and {Petit}, C. and {Baruffolo}, A. and {Feldt}, M. and {Milli}, J. and {Wahhaj}, Z. and {Abe}, L. and {Anselmi}, U. and {Antichi}, J. and {Barette}, R. and {Baudrand}, J. and {Baudoz}, P. and {Bazzon}, A. and {Bernardi}, P. and {Blanchard}, P. and {Brast}, R. and {Bruno}, P. and {Buey}, T. and {Carbillet}, M. and {Carle}, M. and {Cascone}, E. and {Chapron}, F. and {Charton}, J. and {Chauvin}, G. and {Claudi}, R. and {Costille}, A. and {De Caprio}, V. and {de Boer}, J. and {Delboulb{\'e}}, A. and {Desidera}, S. and {Dominik}, C. and {Downing}, M. and {Dupuis}, O. and {Fabron}, C. and {Fantinel}, D. and {Farisato}, G. and {Feautrier}, P. and {Fedrigo}, E. and {Fusco}, T. and {Gigan}, P. and {Ginski}, C. and {Girard}, J. and {Giro}, E. and {Gisler}, D. and {Gluck}, L. and {Gry}, C. and {Henning}, T. and {Hubin}, N. and {Hugot}, E. and {Incorvaia}, S. and {Jaquet}, M. and {Kasper}, M. and {Lagadec}, E. and {Lagrange}, A.-M. and {Le Coroller}, H. and {Le Mignant}, D. and {Le Ruyet}, B. and {Lessio}, G. and {Lizon}, J.-L. and {Llored}, M. and {Lundin}, L. and {Madec}, F. and {Magnard}, Y. and {Marteaud}, M. and {Martinez}, P. and {Maurel}, D. and {M{\'e}nard}, F. and {Mesa}, D. and {M{\"o}ller-Nilsson}, O. and {Moulin}, T. and {Moutou}, C. and {Orign{\'e}}, A. and {Parisot}, J. and {Pavlov}, A. and {Perret}, D. and {Pragt}, J. and {Puget}, P. and {Rabou}, P. and {Ramos}, J. and {Reess}, J.-M. and {Rigal}, F. and {Rochat}, S. and {Roelfsema}, R. and {Rousset}, G. and {Roux}, A. and {Saisse}, M. and {Salasnich}, B. and {Santambrogio}, E. and {Scuderi}, S. and {Segransan}, D. and {Sevin}, A. and {Siebenmorgen}, R. and {Soenke}, C. and {Stadler}, E. and {Suarez}, M. and {Tiph{\`e}ne}, D. and {Turatto}, M. and {Udry}, S. and {Vakili}, F. and {Waters}, L.~B.~F.~M. and {Weber}, L. and {Wildi}, F. and {Zins}, G. and {Zurlo}, A.},
        title = "{SPHERE: the exoplanet imager for the Very Large Telescope}",
      journal = {\aap},
         year = 2019,
        month = nov,
       volume = {631},
          eid = {A155},
        pages = {A155},
          doi = {10.1051/0004-6361/201935251},
archivePrefix = {arXiv},
       eprint = {1902.04080},
 primaryClass = {astro-ph.IM},
       adsurl = {https://ui.adsabs.harvard.edu/abs/2019A&A...631A.155B}
}

@INPROCEEDINGS{Dohlen2008,
       author = {{Dohlen}, Kjetil and {Langlois}, Maud and {Saisse}, Michel and {Hill}, Lucien and {Origne}, Alain and {Jacquet}, Marc and {Fabron}, Christophe and {Blanc}, Jean-Claude and {Llored}, Marc and {Carle}, Michael and {Moutou}, Claire and {Vigan}, Arthur and {Boccaletti}, Anthony and {Carbillet}, Marcel and {Mouillet}, David and {Beuzit}, Jean-Luc},
        title = "{The infra-red dual imaging and spectrograph for SPHERE: design and performance}",
    booktitle = {Ground-based and Airborne Instrumentation for Astronomy II},
         year = 2008,
       editor = {{McLean}, Ian S. and {Casali}, Mark M.},
       series = {Society of Photo-Optical Instrumentation Engineers (SPIE) Conference Series},
       volume = {7014},
        month = jul,
          eid = {70143L},
        pages = {70143L},
          doi = {10.1117/12.789786},
       adsurl = {https://ui.adsabs.harvard.edu/abs/2008SPIE.7014E..3LD}
}

@ARTICLE{vanHolstein2020,
       author = {{van Holstein}, R.~G. and {Girard}, J.~H. and {de Boer}, J. and {Snik}, F. and {Milli}, J. and {Stam}, D.~M. and {Ginski}, C. and {Mouillet}, D. and {Wahhaj}, Z. and {Schmid}, H.~M. and {Keller}, C.~U. and {Langlois}, M. and {Dohlen}, K. and {Vigan}, A. and {Pohl}, A. and {Carbillet}, M. and {Fantinel}, D. and {Maurel}, D. and {Orign{\'e}}, A. and {Petit}, C. and {Ramos}, J. and {Rigal}, F. and {Sevin}, A. and {Boccaletti}, A. and {Le Coroller}, H. and {Dominik}, C. and {Henning}, T. and {Lagadec}, E. and {M{\'e}nard}, F. and {Turatto}, M. and {Udry}, S. and {Chauvin}, G. and {Feldt}, M. and {Beuzit}, J.-L.},
        title = "{Polarimetric imaging mode of VLT/SPHERE/IRDIS. II. Characterization and correction of instrumental polarization effects}",
      journal = {\aap},
         year = 2020,
        month = jan,
       volume = {633},
          eid = {A64},
        pages = {A64},
          doi = {10.1051/0004-6361/201834996},
archivePrefix = {arXiv},
       eprint = {1909.13108},
 primaryClass = {astro-ph.IM},
       adsurl = {https://ui.adsabs.harvard.edu/abs/2020A&A...633A..64V}
}

@MISC{SPHEREManualsESO,
      author = {{ESO}},
      year = 2026,
      title = "{SPHERE instrument documentation and manuals}",
      howpublished = {\url{https://www.eso.org/sci/facilities/paranal/instruments/sphere/doc.html}},
      note = {Accessed: 2026-06-18}
}

@ARTICLE{McCabe2002,
       author = {{McCabe}, C. and {Duch{\^e}ne}, G. and {Ghez}, A.~M.},
        title = "{NICMOS Images of the GG Tauri Circumbinary Disk}",
      journal = {\apj},
         year = 2002,
        month = aug,
       volume = {575},
       number = {2},
        pages = {974-988},
          doi = {10.1086/341479},
archivePrefix = {arXiv},
       eprint = {astro-ph/0204465},
 primaryClass = {astro-ph},
       adsurl = {https://ui.adsabs.harvard.edu/abs/2002ApJ...575..974M}
}

@ARTICLE{Duchene2006,
       author = {{Duch{\^e}ne}, G. and {Beust}, H. and {Adjali}, F. and {Konopacky}, Q.~M. and {Ghez}, A.~M.},
        title = "{Accurate stellar masses in the multiple system T Tauri}",
      journal = {\aap},
         year = 2006,
        month = oct,
       volume = {457},
       number = {1},
        pages = {L9-L12},
          doi = {10.1051/0004-6361:20065917},
archivePrefix = {arXiv},
       eprint = {astro-ph/0608018},
 primaryClass = {astro-ph},
       adsurl = {https://ui.adsabs.harvard.edu/abs/2006A&A...457L...9D}
}

@ARTICLE{Schaefer2020,
       author = {{Schaefer}, G.~H. and {Beck}, Tracy L. and {Prato}, L. and {Simon}, M.},
        title = "{Orbital Motion, Variability, and Masses in the T Tauri Triple System}",
      journal = {\aj},
         year = 2020,
        month = jul,
       volume = {160},
       number = {1},
          eid = {35},
        pages = {35},
          doi = {10.3847/1538-3881/ab93be},
archivePrefix = {arXiv},
       eprint = {2006.03183},
 primaryClass = {astro-ph.SR},
       adsurl = {https://ui.adsabs.harvard.edu/abs/2020AJ....160...35S}
}

@ARTICLE{KrausHillenbrand2009,
       author = {{Kraus}, Adam L. and {Hillenbrand}, Lynne A.},
        title = "{The Coevality of Young Binary Systems}",
      journal = {\apj},
         year = 2009,
        month = oct,
       volume = {704},
       number = {1},
        pages = {531-547},
          doi = {10.1088/0004-637X/704/1/531},
archivePrefix = {arXiv},
       eprint = {0909.0509},
 primaryClass = {astro-ph.SR},
       adsurl = {https://ui.adsabs.harvard.edu/abs/2009ApJ...704..531K}
}

@ARTICLE{KenyonHartmann1995,
       author = {{Kenyon}, Scott J. and {Hartmann}, Lee},
        title = "{Pre-Main-Sequence Evolution in the Taurus-Auriga Molecular Cloud}",
      journal = {\apjs},
         year = 1995,
        month = nov,
       volume = {101},
        pages = {117},
          doi = {10.1086/192235},
       adsurl = {https://ui.adsabs.harvard.edu/abs/1995ApJS..101..117K}
}

@ARTICLE{Ragusa2020,
       author = {{Ragusa}, Enrico and {Alexander}, Richard and {Calcino}, Josh and {Hirsh}, Kieran and {Price}, Daniel J.},
        title = "{The evolution of large cavities and disc eccentricity in circumbinary discs}",
      journal = {\mnras},
         year = 2020,
        month = dec,
       volume = {499},
       number = {3},
        pages = {3362-3380},
          doi = {10.1093/mnras/staa2954},
archivePrefix = {arXiv},
       eprint = {2009.10738},
 primaryClass = {astro-ph.EP},
       adsurl = {https://ui.adsabs.harvard.edu/abs/2020MNRAS.499.3362R}
}

@ARTICLE{Naoz2013,
       author = {{Naoz}, Smadar and {Farr}, Will M. and {Lithwick}, Yoram and {Rasio}, Frederic A. and {Teyssandier}, Jean},
        title = "{Secular dynamics in hierarchical three-body systems}",
      journal = {\mnras},
         year = 2013,
        month = may,
       volume = {431},
       number = {3},
        pages = {2155-2171},
          doi = {10.1093/mnras/stt302},
archivePrefix = {arXiv},
       eprint = {1107.2414},
 primaryClass = {astro-ph.EP},
       adsurl = {https://ui.adsabs.harvard.edu/abs/2013MNRAS.431.2155N}
}

@ARTICLE{Lepp2025,
       author = {{Lepp}, Stephen and {Martin}, Rebecca G. and {Lubow}, Stephen H.},
        title = "{Polar Circumtriple Planets and Disks around Misaligned Hierarchical Triple Stars}",
      journal = {\apj},
         year = 2025,
        month = apr,
       volume = {983},
       number = {2},
          eid = {167},
        pages = {167},
          doi = {10.3847/1538-4357/adc111},
archivePrefix = {arXiv},
       eprint = {2503.06787},
 primaryClass = {astro-ph.EP},
       adsurl = {https://ui.adsabs.harvard.edu/abs/2025ApJ...983..167L}
}

@ARTICLE{Hartigan2003,
       author = {{Hartigan}, Patrick and {Kenyon}, Scott J.},
        title = "{A Spectroscopic Survey of Subarcsecond Binaries in the Taurus-Auriga Dark Cloud with the Hubble Space Telescope}",
      journal = {\apj},
         year = 2003,
        month = jan,
       volume = {583},
       number = {1},
        pages = {334-357},
          doi = {10.1086/345293},
archivePrefix = {arXiv},
       eprint = {astro-ph/0209608},
 primaryClass = {astro-ph},
       adsurl = {https://ui.adsabs.harvard.edu/abs/2003ApJ...583..334H}
}

@ARTICLE{Ghez1995,
       author = {{Ghez}, A.~M. and {Weinberger}, A.~J. and {Neugebauer}, G. and {Matthews}, K. and {McCarthy}, Jr., D.~W.},
        title = "{Speckle Imaging Measurements of the Relative Tangential Velocities of the Components of T Tauri Binary Stars}",
      journal = {\aj},
         year = 1995,
        month = aug,
       volume = {110},
        pages = {753},
          doi = {10.1086/117560},
       adsurl = {https://ui.adsabs.harvard.edu/abs/1995AJ....110..753G}
}

@ARTICLE{Roddier1996,
       author = {{Roddier}, C. and {Roddier}, F. and {Northcott}, M.~J. and {Graves}, J.~E. and {Jim}, K.},
        title = "{Adaptive Optics Imaging of GG Tauri: Optical Detection of the Circumbinary Ring}",
      journal = {\apj},
         year = 1996,
        month = may,
       volume = {463},
        pages = {326},
          doi = {10.1086/177245},
       adsurl = {https://ui.adsabs.harvard.edu/abs/1996ApJ...463..326R}
}

@ARTICLE{Woitas2001,
       author = {{Woitas}, J. and {K{\"o}hler}, R. and {Leinert}, Ch.},
        title = "{Orbital motion in T Tauri binary systems}",
      journal = {\aap},
         year = 2001,
        month = apr,
       volume = {369},
        pages = {249-262},
          doi = {10.1051/0004-6361:20010135},
archivePrefix = {arXiv},
       eprint = {astro-ph/0101450},
 primaryClass = {astro-ph},
       adsurl = {https://ui.adsabs.harvard.edu/abs/2001A&A...369..249W}
}

@ARTICLE{Tamazian2002,
       author = {{Tamazian}, Vakhtang S. and {Docobo}, Jos{\'e} A. and {White}, Russel J. and {Woitas}, Jens},
        title = "{Preliminary Orbits and System Masses for Five Binary T Tauri Stars}",
      journal = {\apj},
         year = 2002,
        month = oct,
       volume = {578},
       number = {2},
        pages = {925-934},
          doi = {10.1086/342621},
       adsurl = {https://ui.adsabs.harvard.edu/abs/2002ApJ...578..925T}
}

@ARTICLE{White2001,
       author = {{White}, R.~J. and {Ghez}, A.~M.},
        title = "{Observational Constraints on the Formation and Evolution of Binary Stars}",
      journal = {\apj},
         year = 2001,
        month = jul,
       volume = {556},
       number = {1},
        pages = {265-295},
          doi = {10.1086/321542},
archivePrefix = {arXiv},
       eprint = {astro-ph/0103098},
 primaryClass = {astro-ph},
       adsurl = {https://ui.adsabs.harvard.edu/abs/2001ApJ...556..265W}
}

@ARTICLE{Gaia2018,
       author = {{Gaia Collaboration} and {Brown}, A.~G.~A. and {Vallenari}, A. and {Prusti}, T. and {de Bruijne}, J.~H.~J. and {Babusiaux}, C. and {Bailer-Jones}, C.~A.~L. and {Biermann}, M. and {Evans}, D.~W. and {Eyer}, L. and {Jansen}, F. and {Jordi}, C. and {Klioner}, S.~A. and {Lammers}, U. and {Lindegren}, L. and {Luri}, X. and {Mignard}, F. and {Panem}, C. and {Pourbaix}, D. and {Randich}, S. and {Sartoretti}, P. and {Siddiqui}, H.~I. and {Soubiran}, C. and {van Leeuwen}, F. and {Walton}, N.~A. and {Arenou}, F. and {Bastian}, U. and {Cropper}, M. and {Drimmel}, R. and {Katz}, D. and {Lattanzi}, M.~G. and {Bakker}, J. and {Cacciari}, C. and {Casta{\~n}eda}, J. and {Chaoul}, L. and {Cheek}, N. and {De Angeli}, F. and {Fabricius}, C. and {Guerra}, R. and {Holl}, B. and {Masana}, E. and {Messineo}, R. and {Mowlavi}, N. and {Nienartowicz}, K. and {Panuzzo}, P. and {Portell}, J. and {Riello}, M. and {Seabroke}, G.~M. and {Tanga}, P. and {Th{\'e}venin}, F. and {Gracia-Abril}, G. and {Comoretto}, G. and {Garcia-Reinaldos}, M. and {Teyssier}, D. and {Altmann}, M. and {Andrae}, R. and {Audard}, M. and {Bellas-Velidis}, I. and {Benson}, K. and {Berthier}, J. and {Blomme}, R. and {Burgess}, P. and {Busso}, G. and {Carry}, B. and {Cellino}, A. and {Clementini}, G. and {Clotet}, M. and {Creevey}, O. and {Davidson}, M. and {De Ridder}, J. and {Delchambre}, L. and {Dell'Oro}, A. and {Ducourant}, C. and {Fern{\'a}ndez-Hern{\'a}ndez}, J. and {Fouesneau}, M. and {Fr{\'e}mat}, Y. and {Galluccio}, L. and {Garc{\'\i}a-Torres}, M. and {Gonz{\'a}lez-N{\'u}{\~n}ez}, J. and {Gonz{\'a}lez-Vidal}, J.~J. and {Gosset}, E. and {Guy}, L.~P. and {Halbwachs}, J.-L. and {Hambly}, N.~C. and {Harrison}, D.~L. and {Hern{\'a}ndez}, J. and {Hestroffer}, D. and {Hodgkin}, S.~T. and {Hutton}, A. and {Jasniewicz}, G. and {Jean-Antoine-Piccolo}, A. and {Jordan}, S. and {Korn}, A.~J. and {Krone-Martins}, A. and {Lanzafame}, A.~C. and {Lebzelter}, T. and {L{\"o}ffler}, W. and {Manteiga}, M. and {Marrese}, P.~M. and {Mart{\'\i}n-Fleitas}, J.~M. and {Moitinho}, A. and {Mora}, A. and {Muinonen}, K. and {Osinde}, J. and {Pancino}, E. and {Pauwels}, T. and {Petit}, J.-M. and {Recio-Blanco}, A. and {Richards}, P.~J. and {Rimoldini}, L. and {Robin}, A.~C. and {Sarro}, L.~M. and {Siopis}, C. and {Smith}, M. and {Sozzetti}, A. and {S{\"u}veges}, M. and {Torra}, J. and {van Reeven}, W. and {Abbas}, U. and {Abreu Aramburu}, A. and {Accart}, S. and {Aerts}, C. and {Altavilla}, G. and {{\'A}lvarez}, M.~A. and {Alvarez}, R. and {Alves}, J. and {Anderson}, R.~I. and {Andrei}, A.~H. and {Anglada Varela}, E. and {Antiche}, E. and {Antoja}, T. and {Arcay}, B. and {Astraatmadja}, T.~L. and {Bach}, N. and {Baker}, S.~G. and {Balaguer-N{\'u}{\~n}ez}, L. and {Balm}, P. and {Barache}, C. and {Barata}, C. and {Barbato}, D. and {Barblan}, F. and {Barklem}, P.~S. and {Barrado}, D. and {Barros}, M. and {Barstow}, M.~A. and {Bartholom{\'e} Mu{\~n}oz}, S. and {Bassilana}, J.-L. and {Becciani}, U. and {Bellazzini}, M. and {Berihuete}, A. and {Bertone}, S. and {Bianchi}, L. and {Bienaym{\'e}}, O. and {Blanco-Cuaresma}, S. and {Boch}, T. and {Boeche}, C. and {Bombrun}, A. and {Borrachero}, R. and {Bossini}, D. and {Bouquillon}, S. and {Bourda}, G. and {Bragaglia}, A. and {Bramante}, L. and {Breddels}, M.~A. and {Bressan}, A. and {Brouillet}, N. and {Br{\"u}semeister}, T. and {Brugaletta}, E. and {Bucciarelli}, B. and {Burlacu}, A. and {Busonero}, D. and {Butkevich}, A.~G. and {Buzzi}, R. and {Caffau}, E. and {Cancelliere}, R. and {Cannizzaro}, G. and {Cantat-Gaudin}, T. and {Carballo}, R. and {Carlucci}, T. and {Carrasco}, J.~M. and {Casamiquela}, L. and {Castellani}, M. and {Castro-Ginard}, A. and {Charlot}, P. and {Chemin}, L. and {Chiavassa}, A. and {Cocozza}, G. and {Costigan}, G. and {Cowell}, S. and {Crifo}, F. and {Crosta}, M. and {Crowley}, C. and {Cuypers}, J. and {Dafonte}, C. and {Damerdji}, Y. and {Dapergolas}, A. and {David}, P. and {David}, M. and {de Laverny}, P. and {De Luise}, F.},
        title = "{Gaia Data Release 2. Summary of the contents and survey properties}",
      journal = {\aap},
         year = 2018,
        month = aug,
       volume = {616},
          eid = {A1},
        pages = {A1},
          doi = {10.1051/0004-6361/201833051},
archivePrefix = {arXiv},
       eprint = {1804.09365},
 primaryClass = {astro-ph.GA},
       adsurl = {https://ui.adsabs.harvard.edu/abs/2018A&A...616A...1G}
}

@ARTICLE{Gaia2016,
       author = {{Gaia Collaboration} and {Prusti}, T. and {de Bruijne}, J.~H.~J. and {Brown}, A.~G.~A. and {Vallenari}, A. and {Babusiaux}, C. and {Bailer-Jones}, C.~A.~L. and {Bastian}, U. and {Biermann}, M. and {Evans}, D.~W. and {Eyer}, L. and {Jansen}, F. and {Jordi}, C. and {Klioner}, S.~A. and {Lammers}, U. and {Lindegren}, L. and {Luri}, X. and {Mignard}, F. and {Milligan}, D.~J. and {Panem}, C. and {Poinsignon}, V. and {Pourbaix}, D. and {Randich}, S. and {Sarri}, G. and {Sartoretti}, P. and {Siddiqui}, H.~I. and {Soubiran}, C. and {Valette}, V. and {van Leeuwen}, F. and {Walton}, N.~A. and {Aerts}, C. and {Arenou}, F. and {Cropper}, M. and {Drimmel}, R. and {H{\o}g}, E. and {Katz}, D. and {Lattanzi}, M.~G. and {O'Mullane}, W. and {Grebel}, E.~K. and {Holland}, A.~D. and {Huc}, C. and {Passot}, X. and {Bramante}, L. and {Cacciari}, C. and {Casta{\~n}eda}, J. and {Chaoul}, L. and {Cheek}, N. and {De Angeli}, F. and {Fabricius}, C. and {Guerra}, R. and {Hern{\'a}ndez}, J. and {Jean-Antoine-Piccolo}, A. and {Masana}, E. and {Messineo}, R. and {Mowlavi}, N. and {Nienartowicz}, K. and {Ord{\'o}{\~n}ez-Blanco}, D. and {Panuzzo}, P. and {Portell}, J. and {Richards}, P.~J. and {Riello}, M. and {Seabroke}, G.~M. and {Tanga}, P. and {Th{\'e}venin}, F. and {Torra}, J. and {Els}, S.~G. and {Gracia-Abril}, G. and {Comoretto}, G. and {Garcia-Reinaldos}, M. and {Lock}, T. and {Mercier}, E. and {Altmann}, M. and {Andrae}, R. and {Astraatmadja}, T.~L. and {Bellas-Velidis}, I. and {Benson}, K. and {Berthier}, J. and {Blomme}, R. and {Busso}, G. and {Carry}, B. and {Cellino}, A. and {Clementini}, G. and {Cowell}, S. and {Creevey}, O. and {Cuypers}, J. and {Davidson}, M. and {De Ridder}, J. and {de Torres}, A. and {Delchambre}, L. and {Dell'Oro}, A. and {Ducourant}, C. and {Fr{\'e}mat}, Y. and {Garc{\'\i}a-Torres}, M. and {Gosset}, E. and {Halbwachs}, J.-L. and {Hambly}, N.~C. and {Harrison}, D.~L. and {Hauser}, M. and {Hestroffer}, D. and {Hodgkin}, S.~T. and {Huckle}, H.~E. and {Hutton}, A. and {Jasniewicz}, G. and {Jordan}, S. and {Kontizas}, M. and {Korn}, A.~J. and {Lanzafame}, A.~C. and {Manteiga}, M. and {Moitinho}, A. and {Muinonen}, K. and {Osinde}, J. and {Pancino}, E. and {Pauwels}, T. and {Petit}, J.-M. and {Recio-Blanco}, A. and {Robin}, A.~C. and {Sarro}, L.~M. and {Siopis}, C. and {Smith}, M. and {Smith}, K.~W. and {Sozzetti}, A. and {Thuillot}, W. and {van Reeven}, W. and {Viala}, Y. and {Abbas}, U. and {Abreu Aramburu}, A. and {Accart}, S. and {Aguado}, J.~J. and {Allan}, P.~M. and {Allasia}, W. and {Altavilla}, G. and {{\'A}lvarez}, M.~A. and {Alves}, J. and {Anderson}, R.~I. and {Andrei}, A.~H. and {Anglada Varela}, E. and {Antiche}, E. and {Antoja}, T. and {Ant{\'o}n}, S. and {Arcay}, B. and {Atzei}, A. and {Ayache}, L. and {Bach}, N. and {Baker}, S.~G. and {Balaguer-N{\'u}{\~n}ez}, L. and {Barache}, C. and {Barata}, C. and {Barbier}, A. and {Barblan}, F. and {Baroni}, M. and {Barrado y Navascu{\'e}s}, D. and {Barros}, M. and {Barstow}, M.~A. and {Becciani}, U. and {Bellazzini}, M. and {Bellei}, G. and {Bello Garc{\'\i}a}, A. and {Belokurov}, V. and {Bendjoya}, P. and {Berihuete}, A. and {Bianchi}, L. and {Bienaym{\'e}}, O. and {Billebaud}, F. and {Blagorodnova}, N. and {Blanco-Cuaresma}, S. and {Boch}, T. and {Bombrun}, A. and {Borrachero}, R. and {Bouquillon}, S. and {Bourda}, G. and {Bouy}, H. and {Bragaglia}, A. and {Breddels}, M.~A. and {Brouillet}, N. and {Br{\"u}semeister}, T. and {Bucciarelli}, B. and {Budnik}, F. and {Burgess}, P. and {Burgon}, R. and {Burlacu}, A. and {Busonero}, D. and {Buzzi}, R. and {Caffau}, E. and {Cambras}, J. and {Campbell}, H. and {Cancelliere}, R. and {Cantat-Gaudin}, T. and {Carlucci}, T. and {Carrasco}, J.~M. and {Castellani}, M. and {Charlot}, P. and {Charnas}, J. and {Charvet}, P. and {Chassat}, F. and {Chiavassa}, A. and {Clotet}, M. and {Cocozza}, G. and {Collins}, R.~S. and {Collins}, P. and {Costigan}, G.},
        title = "{The Gaia mission}",
      journal = {\aap},
         year = 2016,
        month = nov,
       volume = {595},
          eid = {A1},
        pages = {A1},
          doi = {10.1051/0004-6361/201629272},
archivePrefix = {arXiv},
       eprint = {1609.04153},
 primaryClass = {astro-ph.IM},
       adsurl = {https://ui.adsabs.harvard.edu/abs/2016A&A...595A...1G}
}

@ARTICLE{Gaia2023,
       author = {{Gaia Collaboration} and {Vallenari}, A. and {Brown}, A.~G.~A. and {Prusti}, T. and {de Bruijne}, J.~H.~J. and {Arenou}, F. and {Babusiaux}, C. and {Biermann}, M. and {Creevey}, O.~L. and {Ducourant}, C. and {Evans}, D.~W. and {Eyer}, L. and {Guerra}, R. and {Hutton}, A. and {Jordi}, C. and {Klioner}, S.~A. and {Lammers}, U.~L. and {Lindegren}, L. and {Luri}, X. and {Mignard}, F. and {Panem}, C. and {Pourbaix}, D. and {Randich}, S. and {Sartoretti}, P. and {Soubiran}, C. and {Tanga}, P. and {Walton}, N.~A. and {Bailer-Jones}, C.~A.~L. and {Bastian}, U. and {Drimmel}, R. and {Jansen}, F. and {Katz}, D. and {Lattanzi}, M.~G. and {van Leeuwen}, F. and {Bakker}, J. and {Cacciari}, C. and {Casta{\~n}eda}, J. and {De Angeli}, F. and {Fabricius}, C. and {Fouesneau}, M. and {Fr{\'e}mat}, Y. and {Galluccio}, L. and {Guerrier}, A. and {Heiter}, U. and {Masana}, E. and {Messineo}, R. and {Mowlavi}, N. and {Nicolas}, C. and {Nienartowicz}, K. and {Pailler}, F. and {Panuzzo}, P. and {Riclet}, F. and {Roux}, W. and {Seabroke}, G.~M. and {Sordo}, R. and {Th{\'e}venin}, F. and {Gracia-Abril}, G. and {Portell}, J. and {Teyssier}, D. and {Altmann}, M. and {Andrae}, R. and {Audard}, M. and {Bellas-Velidis}, I. and {Benson}, K. and {Berthier}, J. and {Blomme}, R. and {Burgess}, P.~W. and {Busonero}, D. and {Busso}, G. and {C{\'a}novas}, H. and {Carry}, B. and {Cellino}, A. and {Cheek}, N. and {Clementini}, G. and {Damerdji}, Y. and {Davidson}, M. and {de Teodoro}, P. and {Nu{\~n}ez Campos}, M. and {Delchambre}, L. and {Dell'Oro}, A. and {Esquej}, P. and {Fern{\'a}ndez-Hern{\'a}ndez}, J. and {Fraile}, E. and {Garabato}, D. and {Garc{\'\i}a-Lario}, P. and {Gosset}, E. and {Haigron}, R. and {Halbwachs}, J.-L. and {Hambly}, N.~C. and {Harrison}, D.~L. and {Hern{\'a}ndez}, J. and {Hestroffer}, D. and {Hodgkin}, S.~T. and {Holl}, B. and {Jan{\ss}en}, K. and {Jevardat de Fombelle}, G. and {Jordan}, S. and {Krone-Martins}, A. and {Lanzafame}, A.~C. and {L{\"o}ffler}, W. and {Marchal}, O. and {Marrese}, P.~M. and {Moitinho}, A. and {Muinonen}, K. and {Osborne}, P. and {Pancino}, E. and {Pauwels}, T. and {Recio-Blanco}, A. and {Reyl{\'e}}, C. and {Riello}, M. and {Rimoldini}, L. and {Roegiers}, T. and {Rybizki}, J. and {Sarro}, L.~M. and {Siopis}, C. and {Smith}, M. and {Sozzetti}, A. and {Utrilla}, E. and {van Leeuwen}, M. and {Abbas}, U. and {{\'A}brah{\'a}m}, P. and {Abreu Aramburu}, A. and {Aerts}, C. and {Aguado}, J.~J. and {Ajaj}, M. and {Aldea-Montero}, F. and {Altavilla}, G. and {{\'A}lvarez}, M.~A. and {Alves}, J. and {Anders}, F. and {Anderson}, R.~I. and {Anglada Varela}, E. and {Antoja}, T. and {Baines}, D. and {Baker}, S.~G. and {Balaguer-N{\'u}{\~n}ez}, L. and {Balbinot}, E. and {Balog}, Z. and {Barache}, C. and {Barbato}, D. and {Barros}, M. and {Barstow}, M.~A. and {Bartolom{\'e}}, S. and {Bassilana}, J.-L. and {Bauchet}, N. and {Becciani}, U. and {Bellazzini}, M. and {Berihuete}, A. and {Bernet}, M. and {Bertone}, S. and {Bianchi}, L. and {Binnenfeld}, A. and {Blanco-Cuaresma}, S. and {Blazere}, A. and {Boch}, T. and {Bombrun}, A. and {Bossini}, D. and {Bouquillon}, S. and {Bragaglia}, A. and {Bramante}, L. and {Breedt}, E. and {Bressan}, A. and {Brouillet}, N. and {Brugaletta}, E. and {Bucciarelli}, B. and {Burlacu}, A. and {Butkevich}, A.~G. and {Buzzi}, R. and {Caffau}, E. and {Cancelliere}, R. and {Cantat-Gaudin}, T. and {Carballo}, R. and {Carlucci}, T. and {Carnerero}, M.~I. and {Carrasco}, J.~M. and {Casamiquela}, L. and {Castellani}, M. and {Castro-Ginard}, A. and {Chaoul}, L. and {Charlot}, P. and {Chemin}, L. and {Chiaramida}, V. and {Chiavassa}, A. and {Chornay}, N. and {Comoretto}, G. and {Contursi}, G. and {Cooper}, W.~J. and {Cornez}, T. and {Cowell}, S. and {Crifo}, F. and {Cropper}, M. and {Crosta}, M. and {Crowley}, C. and {Dafonte}, C. and {Dapergolas}, A. and {David}, M. and {David}, P. and {de Laverny}, P. and {De Luise}, F. and {De March}, R.},
        title = "{Gaia Data Release 3. Summary of the content and survey properties}",
      journal = {\aap},
         year = 2023,
        month = jun,
       volume = {674},
          eid = {A1},
        pages = {A1},
          doi = {10.1051/0004-6361/202243940},
archivePrefix = {arXiv},
       eprint = {2208.00211},
 primaryClass = {astro-ph.GA},
       adsurl = {https://ui.adsabs.harvard.edu/abs/2023A&A...674A...1G}
}

\begin{appendix}
\onecolumn

\section{Supplementary material}

{\renewcommand{\arraystretch}{1.3}
\begin{table*}[htbp]
\centering
\setlength{\tabcolsep}{8pt}
\caption{Reconstructed $Aa$--$Ab_1$ relative astrometry.}
\label{tab:aa_ab1_astrometry}
\begin{tabular*}{0.8\textwidth}{@{\extracolsep{\fill}}l r@{\hspace{0.5em}$\pm$\hspace{-1.2em}}l r@{\hspace{0.5em}$\pm$\hspace{-1.2em}}l c l@{}}
\hline\hline
Date &
\multicolumn{2}{c}{$\delta$RA [mas]} & \multicolumn{2}{c}{$\delta$Dec [mas]} & $\rho$ & Source \\
\hline
02 Nov. 1990 & -40.48 & 9.01 & -256.15 & 9.99 & 0.04 & \citet{Leinert1993} \\
21 Oct. 1991 & -7.80 & 4.65 & -263.27 & 10.08 & 0.06 & \citet{Ghez1995} \\
26 Dec. 1993 & -15.84 & 9.35 & -258.37 & 10.03 & 0.01 & \citet{Roddier1996} \\
27 Jan. 1994 & 6.89 & 2.78 & -244.60 & 4.07 & -0.07 & \citet{Woitas2001} \\
24 Sep. 1994 & 9.40 & 9.11 & -257.35 & 4.12 & 0.09 & \citet{Ghez1995} \\
18 Oct. 1994 & -8.00 & 2.50 & -241.80 & 3.14 & 0.01 & \citet{Ghez1995} \\
22 Dec. 1994 & 7.28 & 8.41 & -238.90 & 5.10 & 0.05 & \citet{Roddier1996} \\
08 Oct. 1995 & 8.77 & 3.07 & -248.35 & 4.08 & -0.03 & \citet{Woitas2001} \\
29 Sep. 1996 & 15.16 & 2.23 & -247.46 & 4.04 & -0.10 & \citet{Woitas2001} \\
06 Dec. 1996 & 17.80 & 5.55 & -246.08 & 4.64 & 0.03 & \citet{White2001} \\
10 Oct. 1997 & 23.70 & 2.28 & -250.78 & 2.05 & 0.02 & \citet{McCabe2002} \\
16 Nov. 1997 & 25.06 & 2.33 & -249.70 & 4.99 & -0.20 & \citet{Woitas2001} \\
10 Oct. 1998 & 41.22 & 2.47 & -260.94 & 3.99 & -0.19 & \citet{Woitas2001} \\
09 Feb. 2001 & 49.86 & 2.74 & -239.87 & 4.16 & -0.25 & \citet{Tamazian2002} \\
12 Dec. 2002 & 56.32 & 6.45 & -243.30 & 2.58 & 0.57 & \citet{Duchene2004} \\
13 Dec. 2003 & 63.65 & 2.22 & -243.18 & 2.07 & 0.05 & \citet{Koehler2011} \\
20 Nov. 2006 & 89.77 & 2.22 & -239.91 & 2.07 & 0.06 & \citet{Koehler2011} \\
05 Oct. 2009 & 107.80 & 2.77 & -226.24 & 2.09 & 0.08 & \citet{Koehler2011} \\
09 Dec. 2010 & 113.94 & 0.90 & -222.74 & 0.73 & -0.30 & \citet{Toci2024} \\
16 Dec. 2011 & 118.75 & 0.95 & -223.45 & 0.78 & 0.20 & \citet{Toci2024} \\
29 Oct. 2012 & 125.57 & 1.06 & -223.74 & 0.77 & 0.61 & \citet{Toci2024} \\
06 Dec. 2012 & 126.65 & 3.57 & -223.30 & 3.53 & 0.02 & \citet{Toci2024} \\
11 Dec. 2014 & 138.04 & 0.50 & -221.83 & 0.64 & -0.01 & \citet{Toci2024} \\
19 Dec. 2017 & 150.93 & 1.58 & -203.56 & 0.53 & 0.10 & \citet{Toci2024} \\
13 Oct. 2019 & 160.95 & 0.66 & -195.80 & 0.59 & -0.11 & \citet{Toci2024} \\
25 Nov. 2024 & 193.84 & 1.22 & -174.77 & 1.33 & -0.60 & This work \\
\hline
\end{tabular*}
\tablefoot{The reconstructed positions were derived from historical $Aa$--$Ab_\mathrm{pc}$ measurements after accounting for the photocenter offset described in Sect.~\ref{sec:Aa_conversion}. For each epoch, the table lists the relative position in Cartesian coordinates, $\delta$RA and $\delta$Dec, together with their correlation coefficient $\rho$, introduced by the conversion from the polar coordinates, separation and position angle.}
\end{table*}
}

\begin{figure*}[htbp]
\centering
\includegraphics[width=\linewidth]{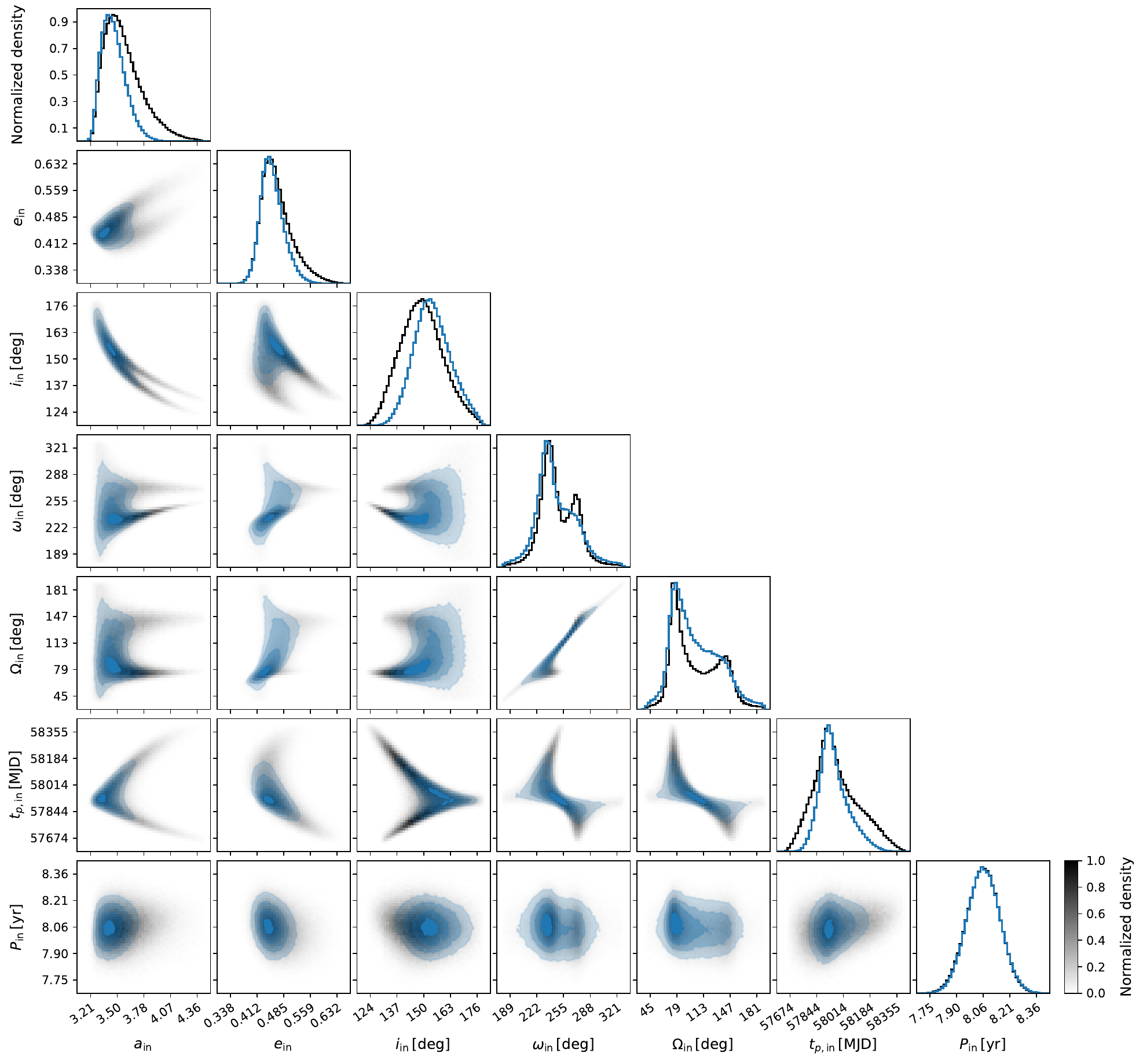}
\caption{Posterior distributions of the inner-orbit parameters for Cases~A and B, shown in black and blue, respectively. This corner plot shows both the 1D marginal distributions and the covariances between parameters. The contours enclose 12\%, 39\%, 68\%, and 87\% of the total posterior probability, with increasing opacity toward higher posterior density. Only one of the two equivalent branches of the astrometric degeneracy is shown. The complementary solution $(\omega+\pi, \Omega+\pi)$ produces the same projected orbit and is therefore equally valid.}
\label{fig:corner_inner}
\end{figure*}

\begin{figure*}[htbp]
\centering
\includegraphics[width=\linewidth]{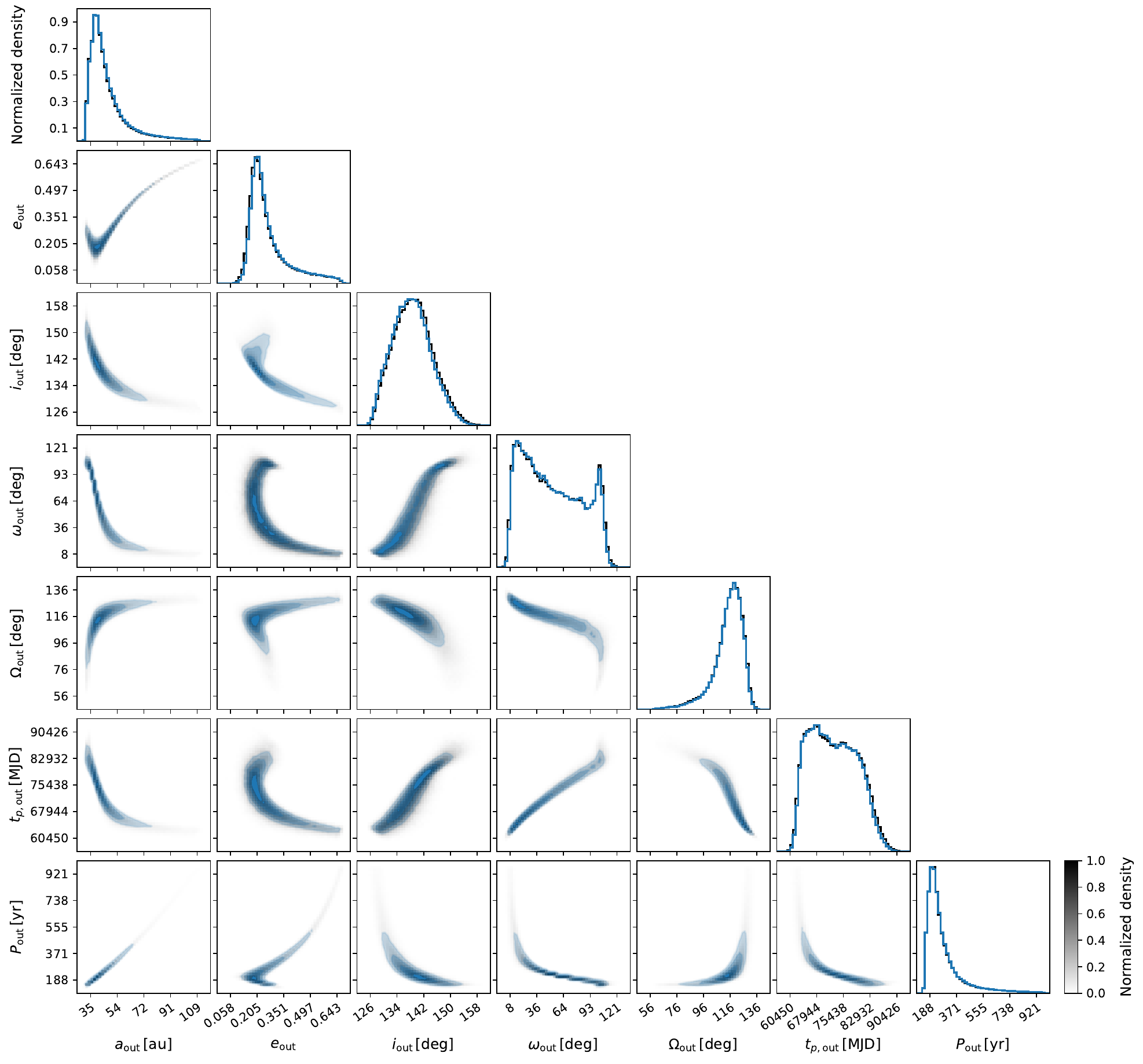}
\caption{Posterior distributions of the outer-orbit parameters for Cases~A and B, shown in black and blue, respectively. Plot conventions are the same as in Fig.~\ref{fig:corner_inner}.}
\label{fig:corner_outer}
\end{figure*}

\end{appendix}

\end{document}